\documentclass{article}

\usepackage{graphicx}
\usepackage{amsmath}
\usepackage{amsfonts}
\usepackage{amssymb}
\usepackage{verbatim}

\renewcommand{\arraystretch}{1.3}

\makeatletter
\newdimen\normalarrayskip              
\newdimen\minarrayskip                 
\normalarrayskip\baselineskip
\minarrayskip\jot
\newif\ifold             \oldtrue            \def\new{\oldfalse}
\def\arraymode{\ifold\relax\else\displaystyle\fi} 
\def\eqnumphantom{\phantom{(\theequation)}}     
\def\@arrayskip{\ifold\baselineskip\z@\lineskip\z@
     \else
     \baselineskip\minarrayskip\lineskip2\minarrayskip\fi}
\def\@arrayclassz{\ifcase \@lastchclass \@acolampacol \or
\@ampacol \or \or \or \@addamp \or
   \@acolampacol \or \@firstampfalse \@acol \fi
\edef\@preamble{\@preamble
  \ifcase \@chnum
     \hfil$\relax\arraymode\@sharp$\hfil
     \or $\relax\arraymode\@sharp$\hfil
     \or \hfil$\relax\arraymode\@sharp$\fi}}
\def\@array[#1]#2{\setbox\@arstrutbox=\hbox{\vrule
     height\arraystretch \ht\strutbox
     depth\arraystretch \dp\strutbox
     width\z@}\@mkpream{#2}\edef\@preamble{\halign
\noexpand\@halignto
\bgroup \tabskip\z@ \@arstrut \@preamble \tabskip\z@ \cr}%
\let\@startpbox\@@startpbox \let\@endpbox\@@endpbox
  \if #1t\vtop \else \if#1b\vbox \else \vcenter \fi\fi
  \bgroup \let\par\relax
  \let\@sharp##\let\protect\relax
  \@arrayskip\@preamble}
\def\eqnarray{\stepcounter{equation}%
              \let\@currentlabel=\theequation
              \global\@eqnswtrue
              \global\@eqcnt\z@
              \tabskip\@centering
              \let\\=\@eqncr

 \halign to \displaywidth\bgroup
    \eqnumphantom\@eqnsel\hskip\@centering
    $\displaystyle \tabskip\z@ {##}$%
    \global\@eqcnt\@ne \hskip 2\arraycolsep
         $\displaystyle\arraymode{##}$\hfil
    \global\@eqcnt\tw@ \hskip 2\arraycolsep
         $\displaystyle\tabskip\z@{##}$\hfil
         \tabskip\@centering
    &{##}\tabskip\z@\cr}
\newfont{\hr}{msbm10}
\newfont{\ams}{msam10}

\voffset= - 1.1in
\hoffset= - 0.9in         

\def\beq{\begin{equation}}
\def\eeq{\end{equation}}
\def\ba{\beq\new\begin{array}{c}}
\def\ea{\end{array}\eeq}
\def\be{\ba}
\def\ee{\ea}

\def\N2{${\cal N}=2$}

\def\1N{${\cal N}=1$}
\def\4N{${\cal N}=4$}
\def\nn{\nonumber}

\def\p{\partial}

\newdimen\linethick  \linethick=0.4pt
\newdimen\hboxitspace    \hboxitspace=5pt
\newdimen\vboxitspace    \vboxitspace=5pt

\def\fr#1{%
\beq\new
\vcenter{
\hrule height\linethick
          \hbox{\vrule width\linethick
                \kern\hboxitspace
                \vbox{\kern\vboxitspace
                      \hbox{$\begin{array}{c}\displaystyle#1
         \end{array}$}%
                      \kern\vboxitspace}%
                \kern\hboxitspace
                \vrule width\linethick}%
          \hrule height\linethick}%
\eeq}

\def\p{\partial}

\def\no{\stackrel{\circ}{_\circ}}
\def\nn{\nonumber}

\def\p{\partial}

\def\langle{\left<}
\def\rangle{\right>}

\title{\vspace{-1cm}{\bf
CFT exercises for the needs of AGT} \vspace{.5cm}}
\author{{\bf Andrei Mironov}\footnote{ {\small {\it
Lebedev Physics Institute} and {\it ITEP, Moscow, Russia}};
mironov@itep.ru; mironov@lpi.ru}, \ {\bf Sergey Mironov}\footnote{
{\small {\it Moscow State University} and {\it ITEP, Moscow,
Russia}}; badzilla@rambler.ru}, \ {\bf Alexei Morozov}\thanks{{\small
{\it ITEP, Moscow, Russia}}; morozov@itep.ru} \ and {\bf Andrey
Morozov}\thanks{{\small {\it Moscow State University} and {\it ITEP,
Moscow, Russia}}; Andrey.Morozov@itep.ru}\date{ }}

\begin{document}

\setcounter{footnote}{3}

\maketitle

\vspace{-4.7cm}

\begin{center}
\hfill FIAN/TD-17/09\\
\hfill ITEP/TH-33/09
\end{center}

\vspace{2.2cm}

\begin{abstract}
An explicit check of the AGT relation between
the $W_N$-symmetry controlled conformal blocks and
$U(N)$ Nekrasov functions requires knowledge of the
Shapovalov matrix and various triple correlators
for $W$-algebra descendants.
We collect simplest expressions of this type
for $N=3$ and for the two lowest descendant levels,
together with the detailed derivations, which can
be now computerized and used in more general studies
of conformal blocks and AGT relations at higher levels.
\end{abstract}

\section{Introduction}

Two-dimensional conformal field theory \cite{Pol}-\cite{BPZ} has
a long and glorious history \cite{Za1}-\cite{DiF}.
It has a number of direct applications
in solid state physics and group theory. As an explicitly solvable
example, it is the basis of our understanding of quantum field
theory. It captures the main algebro-geometric structures, relevant
for description of strings. It serves as a dream-model for any other
branch of theoretical physics, setting the quality level for any
self-consistent description "from the first principles". Still, the
most beautiful parts of conformal field theory remain poorly
demanded by modern theory. A remarkable exception is a recent
suggestion \cite{AGT} to expand the conformal blocks of Liouville
and affine Toda models \cite{Wyl} in series over Nekrasov functions \cite{Nek}.
This AGT relation should be an adequate description for expansion of
Fateev-Dotsenko screening-charge integrals in powers of
$\alpha$-parameters, which is known to produce integrals very much
resembling those which were nicely expanded in \cite{Nek} in sums
over Young diagrams and further related to various issues in
representation theory and Hurwitz-Kontsevich models \cite{HuKo}.
However, the study of the
AGT relations \cite{Wyl,MMMagt,Ga,mmAGT} is surprisingly slowed down
by the lack of the relevant formulas in CFT textbooks, and this is the goal of the
present paper to list some of them.

The main point of our interest here is the conformal block. If
the structure constants of conformal model are known, one can
recursively calculate conformal blocks level by level, exploiting
the conformal symmetry with the standard procedure of
\cite{BPZ,ZaZa,DiF}. Sometimes it is more effective to use recurrent
relations suggested in various forms in \cite{Za1,Za2,ZaZa}. We,
however, follow the third way, close in spirit to that of \cite{Sonoda,MS}.
That is, we construct the conformal block from triple
vertices and the inverse of the Shapovalov matrix. Therefore, our main goal in
this paper is to derive and double-check the simplest expressions
for the Virasoro and $W$-algebra triple correlators (they are frame-boxed
in the text), which are used for constructing the 4-point
conformal blocks, in particular, in AGT studies in \cite{MMMagt} and
especially in \cite{mmAGT}. Note that more triple vertices are needed
for calculation of multi-point
conformal blocks already at level two, but we do not discuss here.

While the Virasoro conformal blocks are well studied, formulas for
the $W_3$-algebra conformal blocks are less available in literature,
moreover, in contrast with the Virasoro case, we are not aware of
any general formulas for triple vertices in this case. Therefore, we
are mostly interested in this latter, more involved case. Still,
since the methods used and a part of answers should be well known to
experts, we organized the text into a simple pedagogical review,
which can be used by newcomers for further advances in the field.

We begin with the model of $2d$ massless free fields as a main
prototype of generic $2d$ conformal theory. All chiral (holomorphic)
correlators in this model can be immediately calculated with the
help of the Wick theorem. Since this is enough for our purposes we
consider only correlators on the sphere, though generalization to
arbitrary Riemann surfaces is straightforward \cite{GMMOS}. Any
general relation between correlators can easily be checked by
explicit calculation of both sides in the free field model. After
that we consider generic decomposition of conformal blocks into
triple vertices {\it of two kinds} and inverse Shapovalov form and
derive various recursive relations between triple vertices with
different number of descendants. Then some of these relations are
explicitly checked in the model of free fields. Finally we make a
summary of recursion relations, needed for the study of AGT
conjecture in \cite{MMMagt,mmAGT}, including partial restriction to
{\it special} states in the case of a theory with $W^{(3)}$ chiral
algebra (of which the model of two free fields is an example).

\section{Ordinary free field with $c=1$}

The massless free field model in $2$ dimensions is defined by the
action $\frac{1}{2}\int \p\phi\bar\p\phi d^2z$, but for our
purposes it is more convenient to define it directly
with the help of the Wick theorem.
We consider holomorphic (chiral) correlators,
and they are defined as sums over all possible pairings
of fields $\phi$ at different points with the simple formulas
for propagators:
\be
<\phi(z_1) \phi(z_2)>\ = \log(z_1-z_2),\\
<\p\phi(z_1) \phi(z_2)>\ = \frac{1}{z_{12}}, \ \ \
<\phi(z_1) \p\phi(z_2)>\ = -\frac{1}{z_{12}},\\
<\p\phi(z_1) \p\phi(z_2)>\ = \frac{1}{z^2_{12}},  \\
\ldots \ee Here $z_{ij} \equiv z_i - z_j$. By definition the {\it
normal ordering} means that all pairings of operators {\it in
between colons} are omitted in application of the Wick theorem.
Average of a normal ordered operator is equal to its constant
($\phi$-independent) item. In particular \be \prod_i
\no e^{\alpha_i\phi(z_i)}\no
\ = \left(\prod_{i<j}
z_{ij}^{\alpha_i\alpha_j}\right) \no\prod_i e^{\alpha_i\phi(z_i)}\no \ \
\ \Longrightarrow\ \ \ \langle \prod_i
\no e^{\alpha_i\phi(z_i)}\no \rangle \ = \left(\prod_{i<j}
z_{ij}^{\alpha_i\alpha_j}\right)
\cdot\delta\left(\sum_i\alpha_i-2Q\right) \label{expocorr} \ee The
delta-function factor at the r.h.s. can be interpreted as arising
from the zero-mode integration in functional integral, however we
simply introduce it as a peculiar {\it selection rule}, specific for
the free field model. Let us begin with the case when $\boxed{Q=0.}$

\begin{figure}[piccont]
\unitlength 1mm 
\linethickness{0.4pt}
\ifx\plotpoint\undefined\newsavebox{\plotpoint}\fi 
\begin{picture}(96.747,35.05)(-35,10)
\put(13.081,28.955){\circle{7.135}}
\put(70.758,28.955){\circle{7.135}}
\put(52.92,28.881){\circle{7.135}}
\put(26.116,12.484){\circle{7.135}}
\put(50.923,12.484){\circle{7.135}}
\put(67.11,12.484){\circle{7.135}}
\put(33.298,28.955){\circle{7.135}}
\put(90.974,28.955){\circle{7.135}}
\put(13.081,28.881){\circle{11.545}}
\put(70.758,28.881){\circle{11.545}}
\put(33.298,28.881){\circle{11.545}}
\put(90.974,28.881){\circle{11.545}}
\put(8.919,35.05){\makebox(0,0)[cc]{$m$}}
\put(66.596,35.05){\makebox(0,0)[cc]{$n$}}
\put(29.136,35.05){\makebox(0,0)[cc]{$n$}}
\put(86.812,35.05){\makebox(0,0)[cc]{$n$}}
\put(13.074,30.969){\makebox(0,0)[cc]{$n$}}
\put(70.751,30.969){\makebox(0,0)[cc]{$m$}}
\put(33.291,30.969){\makebox(0,0)[cc]{$m$}}
\put(90.968,30.969){\makebox(0,0)[cc]{$m$}}
\put(22.298,29.253){\line(1,0){1.189}}
\put(79.974,29.253){\line(1,0){1.189}}
\put(58.216,13.171){\line(1,0){1.189}}
\put(44.386,29.476){\makebox(0,0)[cc]{=}}
\put(35.002,13.183){\makebox(0,0)[cc]{=}}
\put(15.662,13.183){\makebox(0,0)[cc]{=}}
\put(51.136,31.928){\circle{3.006}}
\put(24.332,15.531){\circle{3.006}}
\put(49.139,15.531){\circle{3.006}}
\put(65.326,15.531){\circle{3.006}}
\put(48.014,33.415){\makebox(0,0)[cc]{$m$}}
\put(21.211,17.017){\makebox(0,0)[cc]{$m$}}
\put(46.017,17.017){\makebox(0,0)[cc]{$m$}}
\put(62.204,17.017){\makebox(0,0)[cc]{$n$}}
\put(57.264,32.077){\makebox(0,0)[cc]{$n$}}
\put(30.461,15.68){\makebox(0,0)[cc]{$n$}}
\put(55.267,15.68){\makebox(0,0)[cc]{$n$}}
\put(71.454,15.68){\makebox(0,0)[cc]{$m$}}
\put(60.86,29.116){\makebox(0,0)[cc]{+}}
\qbezier(43.832,18)(40.832,13)(43.832,8)
\qbezier(72.948,18)(75.948,13)(72.948,8)
\put(39.943,13.034){\makebox(0,0)[cc]{$\frac{1}{2}$}}
\end{picture}
\caption{\footnotesize
Calculation of commutator by contours-interchange procedure.}
\label{piccont}
\end{figure}

Then the stress tensor is the operator
\be
T(z) \equiv \frac{1}{2}\no (\p\phi)^2(z)\no
\ee
and it obeys
\be
T(z_1)T(z_2) = \frac{1}{2z_{12}^4}
+ \frac{1}{z_{12}^2}\no \p\phi(z_1)\p\phi(z_2)\no  =
\frac{c}{2z_{12}^4}
+ \frac{2}{z_{12}^2}T(z_2) + \frac{1}{z_{12}}\p T(z_2)
+ \sum_{k\geq 0}\frac{z_{12}^k}{(k+2)!} \no \p^{k+1}\phi\,\p\phi(z_2)\no
\label{TT}
\ee
i.e. the central charge $c$ is equal to 1 in this case.
Its constituents (Virasoro operators $L_k$) act on the exponentials
as follows:
$$
T(z_1)\, \no e^{\alpha\phi(z_2)}\no  \ \equiv
\sum_{k=-\infty}^\infty \frac{1}{z_{12}^{k+2}}
\no \left(L_{k}e^{\alpha\phi}\right)(z_2)\no \
= \frac{\alpha^2/2}{z_{12}^2}\no e^{\alpha\phi(z_2)}\no \ +
\frac{1}{z_{12}}\no \alpha\p\phi(z_1)\, e^{\alpha\phi(z_2)}\no \ +\
\no T(z_1) e^{\alpha\phi(z_2)}\no
\ =
$$
\vspace{-0.5cm}
\be
= \frac{\alpha^2}{2z_{12}^2}\no e^{\alpha\phi(z_2)}\!\no  +
\frac{1}{z_{12}}\no \!\left(\p e^{\alpha\phi}\right)(z_2)\!\no  +
\no \!\left(\frac{(\p\phi)^2}{2} + \alpha\p^2\phi\right)\!
e^{\alpha\phi}(z_2)\no
+ \sum_{k>0} z_{12}^k
\no \!\!\left(\frac{\p^k T}{k!} + \frac{\alpha\,\p^{k+2}\phi}{(k+1)!}\right)
e^{\alpha\phi}(z_2)\no
\label{Tonexp}
\ee
As a consequence of (\ref{TT}), Virasoro operators form the
{\it Virasoro Lie algebra}, with commutation relations
\be
\left[L_m,L_n\right] = \frac{m(m^2-1)}{12}\delta_{n+m,0} +
(m-n)L_{n+m}
\label{virc}
\ee
It can be derived as follows: if one acts on operators at some point $z_3$,
then, by definition of $L_k$ in (\ref{Tonexp})
\be
\left[L_m,L_n\right] = \oint_{z_2+z_3} z_{13}^{m+1} T(z_1)dz_1
\oint_{z_3}z_{23}^{n+1} T(z_2)dz_2 -
\oint_{z_2+z_3} z_{13}^{n+1} T(z_1)dz_1
\oint_{z_3}z_{23}^{m+1} T(z_2)dz_2
\ee
where the integration contour for the second-acting operator
encircles that for the first-acting one. If now pushing one
contour through the other in the first item, one gets (see Fig.\ref{piccont})
\be
\left[L_m,L_n\right] =
\left(\oint_{z_3} z_{13}^{m+1} T(z_1)dz_1
\oint_{z_1+z_3}z_{23}^{n+1} T(z_2)dz_2 -
\oint_{z_2+z_3} z_{13}^{n+1} T(z_1)dz_1
\oint_{z_3}z_{23}^{m+1} T(z_2)dz_2\right) +
\\
+ \oint_{z_3}z_{23}^{n+1} T(z_2)dz_2 \oint_{z_2} dz_1
z_{13}^{m+1} T(z_1)dz_1
 = \frac{1}{2} \oint_{z_3} dz_2\oint_{z_2} dz_1
 \left(z_{13}^{m+1}z_{23}^{n+1}-z_{13}^{n+1}z_{23}^{m+1}\right)
T(z_1)T(z_2) =
\\
= \frac{1}{2}\oint_{z_3} dz_2 \oint_{z_2} dz_1
\left(z_{13}^{m+1}z_{23}^{n+1}-z_{13}^{n+1}z_{23}^{m+1}\right)
\left(\frac{1}{2z_{12}^4}
+ \frac{2T(z_2)}{z_{12}^2} + \frac{\p T(z_2)}{z_{12}}\right)
\ee
where the very first bracket vanishes
(it is enough to interchange notation for integration variables
$z_1\leftrightarrow z_2$ in one of the items to see this),
while the second integral picks up only contributions from
the terms of (\ref{TT}) which are singular in $z_{12}$.
It is more convenient to rewrite it as a one-half of the
antisymmetrized expression.
The r.h.s. of (\ref{virc}) is now an immediate result of
$z_1$ integration.

\section{Free field with $c\neq 1$}

A slight modification of the free-field theory allows one to
make $Q\neq 0$ in (\ref{expocorr}).
It involves a deformation of the stress tensor
\be
T(z) \equiv \frac{1}{2}\no (\p\phi)^2(z)\no  + Q\p^2\phi(z)
\ee
which now obeys
\be
T(z_1)T(z_2) = \frac{1-12Q^2}{2}\cdot\frac{1}{z_{12}^4}
+2Q\frac{\p\phi(z_1)-\p\phi(z_2)}{z_{12}^3}
+ \frac{1}{z_{12}^2}\no \p\phi(z_1)\p\phi(z_2)\no  = \\ =
\frac{1}{2}\cdot\frac{1}{z_{12}^4}
+ \frac{2}{z_{12}^2}T(z_2) + \frac{1}{z_{12}}\p T(z_2)
+ \sum_{k\geq 0}z_{12}^k\left( \frac{1}{(k+2)!} \no \p^{k+1}\phi\,\p\phi(z_2)\no
+ \frac{2Q}{(k+3)!}\p^{k+4}\phi\right)
\ee
and its action on the exponentials is also modified:
\be
T(z_1)\, \no e^{\alpha\phi(z_2)}\no  \
= \frac{\alpha(\alpha-2Q)}{2z_{12}^2}\no e^{\alpha\phi(z_2)}\no \ +
\frac{1}{z_{12}}\no \alpha\p\phi(z_1)\, e^{\alpha\phi(z_2)}\no \ +\
\no T(z_1) e^{\alpha\phi(z_2)}\no
\ =
\\
= \frac{(\alpha-Q)^2-Q^2}{2z_{12}^2}\no e^{\alpha\phi(z_2)}\no \ +
\frac{1}{z_{12}}\no \left(\p e^{\alpha\phi}\right)(z_2)\no \ +\
\no \left(\frac{(\p\phi)^2}{2} + (\alpha+Q)\p^2\phi\right)
e^{\alpha\phi}(z_2)\no +\\
+ \sum_{k>0} z_{12}^k
\no \left(\frac{\p^k T}{k!} + \frac{\alpha \p^{k+2}\phi}{(k+1)!}\right)
e^{\alpha\phi}(z_2)\no
\ee
Relation between the shifts of the selection rule
\be
\sum_i \alpha_i = 2Q
\ee
by the "vacuum charge" $2Q$ and the "improvement" of the stress
tensor is dictated by the requirement that the double correlator
\be
\langle \no e^{\alpha_1\phi(z_1)}\no \ \no e^{\alpha_2\phi(z_2)}\no  \rangle\
= z_{12}^{\alpha_1\alpha_2} = {z_{12}^{-2\Delta_{\alpha_1}}}
\ee
is expressed in terms of the common dimension $\Delta_{\alpha_1} =
\frac{(\alpha_1-Q)^2 - Q^2}{2} =  \frac{(\alpha_2-Q)^2 - Q^2}{2}
= \Delta_{\alpha_2}$ of the two exponential operators.
Note that the vacuum charge operator $\no e^{-2Q\phi}\no $
has a non-vanishing dimension.

Virasoro algebra is also deformed to
\be
\left[L_m,L_n\right] = \frac{1-12Q^2}{12}m(m^2-1)\delta_{m+n,0} +
(m-n)L_{n+m}
\label{vircQ}
\ee

Unfortunately, there is no conventional choice of normalizations
in the free field models of CFT. The two popular alternatives
to above version include multiplying by $i$:
$\phi\longrightarrow i\phi$, $\alpha\longrightarrow i\alpha$,
and/or multiplication by $\sqrt{2}$:
$\alpha\longrightarrow\sqrt{2}\alpha$ and
$Q\longrightarrow \sqrt{2}Q$
($\phi$ is left intact in this case, and a peculiar parameter
$D \longrightarrow 8D$ below).
In CFT studies {\it per se} our present normalization is
more convenient, because it allows one to avoid extra factor of $\sqrt{2}$
in most formulas. However, physically normalization {\it with}
roots is more relevant, and it is usually used in application
of CFT to other branches of theory. In particular, with
our {\it present} normalization a factor of $\sqrt{2}$
would appear in the AGT relation, $Q = \epsilon\sqrt{2}$,
hence the {\it other} choice made in the AGT-oriented papers
\cite{MMMagt} and \cite{mmAGT}. Instead, there will be
$\sqrt{2}$ in exponentials for the primary field.

For good or for bad, in this paper we stay with the choice,
described by the formulas in this section.

\section{Decomposition of correlators in the free-field model}

A four-point correlator can be evaluated in two different ways:
either directly or by double application of operator expansion.
For example,
\be
\langle \p\phi(z_1)\ \p\phi(z_2)\ \p\phi(z_3)\ \p\phi(z_4)\rangle\
= \frac{1}{z_{12}^2z_{34}^2} + \frac{1}{z_{13}^2z_{24}^2} +
\frac{1}{z_{14}^2z_{23}^2} \ \stackrel{{\rm Wick\ theorem}}{=}\ \\ =
\frac{1}{z_{12}^2z_{34}^2} + \frac{1}{z_{24}^4}\left(
\frac{1}{\left(1 + \frac{z_{12}}{z_{24}} - \frac{z_{34}}{z_{24}}\right)^2} +
\frac{1}{\left(1 + \frac{z_{12}}{z_{24}}\right)^2
\left(1 - \frac{z_{34}}{z_{24}}\right)^2}
\right) = \\
= \frac{1}{z_{12}^2z_{34}^2} + \frac{1}{z_{24}^4}\left(2\ \ \
-\frac{4z_{12}}{z_{24}} + \frac{4z_{34}}{z_{24}} \ \ \
+\frac{6z_{12}^2}{z_{24}^2} - \frac{10z_{12}z_{34}}{z_{24}^2}
+ \frac{6z_{34}^2}{z_{12}^2} + \ldots
\right)
\ee
At the same time
\be
\langle \p\phi(z_1)\ \p\phi(z_2)\ \p\phi(z_3)\ \p\phi(z_4)\rangle\
= \langle \Big(\p\phi(z_1)\ \p\phi(z_2)\Big)\
\Big(\p\phi(z_3)\ \p\phi(z_4)\Big)\rangle\ =\\
= \langle \left(\frac{1}{z_{12}^2}\ + \ \no \p\phi(z_1)\ \p\phi(z_2)\no \right)\
\left(\frac{1}{z_{34}^2}\ + \ \no \p\phi(z_3)\ \p\phi(z_4)\no \right)\rangle\
= \\ = \frac{1}{z_{12}^2z_{34}^2} +
\sum_{k,l\geq 0}\frac{z_{12}^kz_{34}^l}{k!l!}
\langle
\no \p^{k+1}\phi\p\phi(z_2)\no \ \no \p^{l+1}\phi\p\phi(z_4)\no \rangle
\ \stackrel{{\rm Wick\ theorem}}{=}\\
= \frac{1}{z_{12}^2z_{34}^2} +
\sum_{k,l\geq 0}\frac{z_{12}^kz_{34}^l}{k!l!}
\frac{(k+l+1)!+(k+1)!(l+1)!}{z_{24}^{2+k+l}}
\ee
Similarly
\be
\langle \no e^{\alpha_1\phi(z_1)}\no  \ \no e^{\alpha_2\phi(z_2)}\no \
\no e^{\alpha_3\phi(z_3)}\no \ \no e^{\alpha_4\phi(z_4)}\no \rangle \ =
z_{12}^{\alpha_1\alpha_2}z_{13}^{\alpha_1\alpha_3}
z_{14}^{\alpha_1\alpha_4}z_{23}^{\alpha_2\alpha_3}
z_{24}^{\alpha_2\alpha_4}z_{34}^{\alpha_3\alpha_4} =
\\
= z_{12}^{\alpha_1\alpha_2}z_{34}^{\alpha_3\alpha_4}
z_{24}^{(\alpha_1+\alpha_2)(\alpha_3+\alpha_4)}
\left(1+\frac{z_{12}}{z_{24}}\right)^{\alpha_1\alpha_4}
\left(1-\frac{z_{34}}{z_{24}}\right)^{\alpha_3\alpha_4}
\left(1+\frac{z_{12}}{z_{24}}
-\frac{z_{34}}{z_{24}}\right)^{\alpha_1\alpha_3} = \\
= z_{12}^{\alpha_1\alpha_2}z_{34}^{\alpha_3\alpha_4}
z_{24}^{(\alpha_1+\alpha_2)(\alpha_3+\alpha_4)}\left(1\ \ \ +
\alpha_1(\alpha_3+\alpha_4)\frac{z_{12}}{z_{24}} -
\alpha_3(\alpha_1+\alpha_2)\frac{z_{34}}{z_{24}} \ \ \ +
\ldots\right)
\ee
and
\be
\langle \no e^{\alpha_1\phi(z_1)}\no  \ \no e^{\alpha_2\phi(z_2)}\no \
\no e^{\alpha_3\phi(z_3)}\no \ \no e^{\alpha_4\phi(z_4)}\no \rangle \ =
z_{12}^{\alpha_1\alpha_2}z_{34}^{\alpha_3\alpha_4}
\langle \no e^{\alpha_1\phi(z_1)+\alpha_2\phi(z_2)}\no \
\no e^{\alpha_3\phi(z_3)+\alpha_4\phi(z_4)}\no \rangle \ =\\
= z_{12}^{\alpha_1\alpha_2}z_{34}^{\alpha_3\alpha_4}
\langle \no
\Big(1+z_{12}\alpha_1\p\phi + \frac{z_{12}^2}{2}\left((\alpha_1\p\phi)^2
+ \alpha_1\p^2\phi\right)+ \ldots\Big)\,e^{(\alpha_1+\alpha_2)\phi}(z_2)\no
\right.\\ \left.
\no \Big(1+z_{34}\alpha_3\p\phi + \frac{z_{34}^2}{2}\left((\alpha_3\p\phi)^2
+ \alpha_3\p^2\phi\right)+ \ldots\Big)\,e^{(\alpha_3+\alpha_4)\phi}(z_4)\no \rangle
= \\ =
z_{12}^{\alpha_1\alpha_2}z_{34}^{\alpha_3\alpha_4}
z_{24}^{(\alpha_1+\alpha_2)(\alpha_3+\alpha_4)}\left(1\ +
\alpha_1(\alpha_3+\alpha_4)\frac{z_{12}}{z_{24}} -
\alpha_3(\alpha_1+\alpha_2)\frac{z_{34}}{z_{24}}\ +
\ldots\right)
\ee

In generic CFT model there is no direct way to find the
multi-point correlators, thus only the second path remains:
decomposition by iterations of operator expansions.
This procedure expresses an arbitrary correlator through the structure
constants of operator expansion and pair correlators,
or, which is the same, through triple and pair correlators.
However in this way correlators are represented by infinite sums
over intermediate states, thus
the knowledge of generic formulas for triple and
pair correlators is desirable.

\section{Decomposition of the four-point function. General case}

Holomorphic conformal field theory is a straightforward
generalization of the free field model, where the Wick theorem is
substituted by the operator product expansion and a selection rule is
released. The notion of correlators is substituted by two: one
linear and one bilinear form on the space of operators. Virasoro
algebra is imposed as a special Hermiticity requirement on the
bilinear form. In full conformal theory there are also additional
rules, allowing to glue holomorphic and antiholomorphic parts into
globally define modular invariant correlators, but we do not address
this artful and sophisticated part of the story in the present
text.

\bigskip

Operator product expansion is a product in an associative ring of vertex operators
\be
V_1(z_1)V_2(z_2) = \sum_{\check\beta} \frac{C^{\check\beta}_{12}}
{z_{12}^{\Delta_1+\Delta_2-\Delta_{\check\beta}}}V_{\check\beta}(z_2)
\label{ope}
\ee
Then application of the linear form, which we refer to as a
holomorphic correlator and denote by angular brackets, gives
\be
\langle V_1(z_1)V_2(z_2) V_{\check\gamma}(z_4)\rangle =
\sum_{\check\beta} \frac{C^{\check\beta}_{12}}
{z_{12}^{\Delta_1+\Delta_2-\Delta_{\check\beta}}}
\langle V_{\check\beta}(z_2) V_{\check\gamma}(z_4)\rangle
\label{3p2p}
\ee
and
\be
\langle V_1(z_1)V_2(z_2) V_3(z_3)V_4(z_4)\rangle =
\langle \Big(V_1(z_1)V_2(z_2)\Big)\,\Big(V_3(z_3)V_4(z_4)\Big)\rangle
= \sum_{\check\beta,\check\gamma}
\frac{C^{\check\beta}_{12}C^{\check\gamma}_{34}}
{z_{12}^{\Delta_1+\Delta_2-\Delta_{\check\beta}}
z_{34}^{\Delta_3+\Delta_4-\Delta_{\check\gamma}}}
\langle V_{\check\beta}(z_2)V_{\check\gamma}(z_4)\rangle
\ee
The first of these relations is manifestly given by the
holomorphic part of correlator in full conformal field theory
and can be used to express
the structure constants $C$ through triple and pair correlators
(hereafter, for the sake of brevity,
we call holomorphic correlators merely as correlators),
which can then be substituted into the second relation
to provide a desired formula for the four-point correlator.

Alternative representation arises if one stops at decomposing only the
product of the first two operators:
\be
\langle V_1(z_1)V_2(z_2) V_3(z_3)V_4(z_4)\rangle =
\langle \Big(V_1(z_1)V_2(z_2)\Big)\,V_3(z_3)V_4(z_4)\rangle
= \frac{1}{z_{12}^{\Delta_1+\Delta_2}}\sum_{\check\beta}
z_{12}^{\Delta_{\check\beta}}C^{\check\beta}_{12}
\langle V_{\check\beta}(z_2)V_3(z_3)V_4(z_4)\rangle
\label{4p23p}
\ee
This expression can be considerably simplified by an artful choice
of points $z_i$.
Putting $z_1=1$, $z_2=0$ and $z_4=\infty$,
we get rid of the $z$-dependence in (\ref{3p2p}):
\be
\langle V_1(1)V_2(0) V_{\check\gamma}(\infty)\rangle =
\sum_{\check\alpha} C^{\check\beta}_{12}
\langle V_{\check\beta}(0) V_{\check\gamma}(\infty)\rangle
\label{3p2pinf}
\ee
Similarly, putting $z_1=x$, $z_2=0$ and $z_3=1$ and $z_4=\infty$,
we obtain from (\ref{4p23p}):\footnote{Note that the correlators
at both sides of (\ref{4p23pinf}) behave in the same way, as
$\Lambda^{-2\Delta_4}$, when the argument $\Lambda$ of
$V_4(\Lambda)$ tends to infinity.
One can multiply both sides of the equation by $\Lambda^{2\Delta_4}$
and then take the limit, to make correlators well defined.
The same concerns (\ref{3p2pinf}), only the correction
factor is $\Lambda^{2\Delta_{\check\gamma}}$.
}
\be
\langle V_1(x)V_2(0) V_3(1)V_4(\infty)\rangle
= x^{-\Delta_1-\Delta_2} \sum_{\check\beta} x^{\Delta_{\check\beta}}
C_{12}^{\check\beta}
\langle V_{\check\beta}(0)V_3(1)V_4(\infty)\rangle
\label{4p23pinf}
\ee

Let us now introduce a scalar product in the space of vertex
operators, and define the {\it Shapovalov matrix}
\be
H_{\check\alpha\check\beta} \equiv \
\langle V_{\check\alpha}| V_{\check\beta}\rangle
\label{Shama}
\ee
To make it well defined we assume that operators in the product
are taken at a certain point, which we put at $z=0$,
then $H_{\check\alpha\check\beta}$ is just a $c$-number matrix.
We specify other requirements for this scalar product later,
in s.\ref{vir}, in particular they would imply that
Shapovalov matrix has a block-diagonal form,
\be
\Delta_{\check\alpha} \neq \Delta_{\check\beta}
\ \ \Longrightarrow\ \ H_{\check\alpha\check\beta} = 0
\label{orthoShap}
\ee
The blocks of the Shapovalov matrix are
{\it often} finite-dimensional,
then the matrix can be  easily inverted.
If one takes the scalar product of (\ref{ope})
at $z_1=1$ and $z_2=0$ with $V_{\check\alpha}$,
\be
\Bar\Gamma_{12;\check\alpha}\ \equiv \
\langle V_{\check\alpha}|V_1(1) V_2(0)\rangle
\ \stackrel{(\ref{ope})}{=}\
\sum_{\check\beta} C^{\check\beta}_{12}
\langle V_{\check\alpha}| V_{\check\beta}\rangle =
\sum_{\check\beta} C^{\check\beta}_{12} H_{\check\alpha\check\beta}
\label{barGC}
\ee
this allows us to express the structure constants in the form
\be
\boxed{
C^{\check\beta}_{12} =
\sum_{\check\alpha}
\bar\Gamma_{12;\check\alpha}(H^{-1})^{\check\alpha\check\beta}
}
\label{CvsG}
\ee
If one now introduces a similar notation for the 3-point function
in (\ref{3p2pinf}),
\be
\Gamma_{\check\beta 34} \equiv
\langle V_{\check\beta}(0)V_3(1) V_{4}(\infty)\rangle
\ee
then one obtains from (\ref{4p23pinf})
\be
\boxed{
K_{1234}(x) \equiv \langle V_1(x)V_2(0) V_3(1)V_4(\infty)\rangle
= x^{-(\Delta_1+\Delta_2)}
\sum_{\check\alpha,\check\beta} x^{\Delta_{\check\alpha}}
\bar\Gamma_{12;\check\alpha}(H^{-1})^{\check\alpha\check\beta}
\Gamma_{\check\beta 34}
}
\label{KvsH}
\ee

Note that, though denoted similarly, $\bar\Gamma$ and $\Gamma$
are defined in absolutely different ways and
have not much  to do with each other, at least, {\it a priori}.
However, if Virasoro symmetry is taken into account,
these two 3-point functions become rather similar.

\section{Virasoro representations \label{vir}}

The Virasoro symmetry of conformal field theory allows one to
classify vertex operators in terms of the Virasoro representation
theory. Each $V_{\check\alpha}$ belongs to some Verma module,
i.e. is obtained by the action of Virasoro operators $L_n$ with
$n<0$ on the highest weight, called primary field in the context
of CFT:
\be
V_{\check\alpha} = \ldots L_{-k_2}\ldots L_{-k_1} V_\alpha =
\prod_k^{\leftarrow}L_{-k}^{n_k} V_\alpha =
L_{-Y}V_\alpha
\ee
where $Y$ is a Young diagram or an ordered integer partition
$Y = \{k_1\geq k_2\geq\ldots\}$ of size $|Y|=k_1+k_2+\ldots$.
Accordingly, $\check\alpha = (\alpha,Y)$ where $\alpha$
labels different primary fields (primaries) and $Y$ -- different
descendants.
The Virasoro symmetry fixes completely all the $Y$ dependence of
correlators. Different conformal models differ by sets
of primaries and by way the correlators depend on $\alpha$.
A big part of conformal field theory studies is devoted to degenerations
of Verma modules, that is, the cases when there exist null-vectors (which
is a feature of {\it rational} conformal models).
We do not touch these issues in the paper.

The main requirement to the scalar product (\ref{Shama})
is that it is consistent with the Virasoro symmetry in the
following sense:
\be
\langle L_{-n} V_{\check\alpha} | V_{\check\beta} \rangle\
= \ \langle V_{\check\alpha} | L_n V_{\check\beta} \rangle\
\label{Hermi}
\ee
i.e. Virasoro operators are Hermitean.
Accordingly, for the $\bar\Gamma$ function one has
\be
\langle L_{-n} V_{\check\alpha}\ |\ V_1(1)V_2(0) \rangle\
=\ \langle V_{\check\alpha}\Big|
\oint_{0+1} x^{n+s-1} dx \ T(x)V_1(1)V_2(0) \rangle\ =\\
= \oint_1 \frac{x^{n+s-1}dx}{(x-1)^{k+s}}
\langle V_{\check\alpha}\ |\ (L_k V_1)(1)\ V_2(0)\rangle\
+ \oint_0 \frac{x^{n+s-1}dx}{x^{k+s}}
\langle V_{\check\alpha}\ |\ V_1(1)\ (L_k V_2)(0)\rangle
\label{barGrel}
\ee
where integration contours encircles both points $z=0$ and $z=1$
and $s=2$ for the Virasoro algebra.
Using operator product expansion of the stress tensor and
vertex operators one can now express $\Gamma_{12;(\alpha,Y)}$
through $\Gamma_{12;\alpha}$.

In a similar way, one can express $\Gamma_{(\beta,Y)34}$
through $\Gamma_{\beta34}$, using the property that, if inserted into
a 3-point function, the stress tensor has singularities
only in the vicinity of the three vertex operators:
\be
\langle (L_{-n}V_{\check\alpha})(0)\ V_3(1)V_4(\infty) \rangle\ =
\oint_0 \frac{dx}{x^{n-s+1}}
\langle T(x) V_{\check\alpha}(0) V_3(1)V_4(\infty) \rangle\ = \\ =
-\oint_1 \frac{dx}{x^{n-s+1}(x-1)^{k+s}}
\langle V_{\check\alpha}(0)\ (L_kV_3)(1)\ V_4(\infty) \rangle\
+ (-)^s \oint_\infty \frac{x^{k-s}dx}{x^{n-s+1}}
\langle V_{\check\alpha}(0)\ V_3(1)\ (L_kV_4)(\infty) \rangle\
\label{Grel}
\ee
The last integrals both in (\ref{barGrel}) and (\ref{Grel})
pick up the contributions from $k=n$ only,
the first integrals can contain several terms.
If $V_1,V_2,V_3$ and $V_4$ are primaries
as we assume everywhere below in this text, then
\be
\bar\Gamma_{12;(\alpha,Y)} = \bar\gamma_{12;\alpha}(Y) C_{12;\alpha},\\
\Gamma_{(\beta,Y)34} = \gamma_{\beta 34}(Y) C_{\beta34},
\label{gammadef}
\ee
where $C_{12;\alpha}\equiv \bar\Gamma_{12;(\alpha,0)}=C^\beta_{12}H_{\alpha\beta}$,
and similarly for $C_{\beta34}$.
The Shapovalov matrix is usually defined so that
$H_{\alpha\beta}=\delta_{\alpha\beta}$,
then $C_{12;\alpha}=C^\alpha_{12}$ (it is also usually put equal to $C_{\alpha 12}$).

In fact, {\it for the Virasoro algebra}
the two 3-point functions $\gamma$ are well known to be
equivalent,
\be
\bar\gamma^{Vir}_{12;\alpha}(Y) = \gamma^{Vir}_{\alpha12}(Y)
\label{barGGvir}
\ee
but this property does {\it not} persist for more complicated
chiral algebras, see below.

For two primaries $V_\alpha$ and $V_\beta$, the Shapovalov matrix
\be
H_{(\alpha,Y),(\beta,Y')} =
\langle L_{-Y}V_\alpha | L_{-Y'}V_\beta \rangle
= \langle V_\alpha | L_YL_{-Y'}V_\beta \rangle
=  \delta_{|Y|,|Y'|} Q_{\alpha\beta}(Y,Y')  H_{\alpha\beta}
\ee
vanishes for $|Y|\neq |Y'|$
because primaries are annihilated by positive Virasoro generators,
\be
L_nV_\alpha = 0\ \ \ \ {\rm for}\ n>0
\ee
This $\delta$-factor is an exact version of (\ref{orthoShap})
which we use in this paper.

In result, we get for the holomorphic  4-point function(\ref{KvsH})
\be
K_{1234}(x) \equiv \langle V_1(x)V_2(0) V_3(1)V_4(\infty)\rangle
= x^{-(\Delta_1+\Delta_2)}
\sum_{\check\alpha,\check\beta} x^{\Delta_{\check\alpha}}
\bar\Gamma_{12;\check\alpha}(H^{-1})^{\check\alpha\check\beta}
\Gamma_{\check\beta 34} =\\
= x^{-(\Delta_1+\Delta_2)}
\sum_{\alpha,\beta} x^{\Delta_\alpha}
\Big(C_{12;\alpha}H^{-1}_{\alpha\beta}C_{\beta34}\Big)
 B_{1234}^{\alpha\beta}(Y|x),
\label{KvsB}
\ee
where {\it the conformal block}
\be
\boxed{
B_{1234}^{\alpha\beta}(Y|x) =
\sum_{|Y|=|Y'|} x^{|Y|} \bar\gamma_{12;\alpha}(Y)
Q^{-1}_{\alpha\beta}(Y,Y')\gamma_{\beta34}(Y')
}
\label{BvsGQ}
\ee

\section{Extended conformal algebra}

Some models which have (in)finitely many Virasoro primaries
can possess extended loop symmetries which control,
partly or fully, the dependence on (in)finitely many remaining
indices $\alpha$ that labels Virasoro primaries.
Such a symmetry is called extended conformal or chiral algebra,
and Virasoro is always a part of it, but the conformal algebra
can be larger. A well known example is the current algebra
in WZNW model, but we need something a little more involved, since, e.g.,
the theory of $r$ free fields is not fully controlled by
Virasoro symmetry, the relevant chiral algebra is $W_{r+1}$.

Given a chiral algebra, one can introduce the corresponding
primaries. A single Verma module of extended chiral algebra
may contain infinitely many Virasoro primaries and the new
conformal blocks are then sums of infinitely many Virasoro
conformal blocks.\footnote{
In fact, in most conformal models, physical correlators
are {\it sums} of bilinear combinations
of holomorphic and antiholomorphic conformal blocks, i.e.
there is no one-to-one correspondence between the conformal block and
the correlator. This means the sum of holomorphic conformal blocks
makes no sense.
An important story about relation between holomorphic and
bilinear sums is provided by the theory of NSR
and heterotic superstrings and GSO projection \cite{NSR,NSRrev}.
}
Elements of Verma module are now labeled by generalized
Young diagrams ${\cal Y}$, which depend on particular
choice of the algebra. Many formulas from s.\ref{vir})
remain the same, if operators $L$ are substituted by the
generators of the chiral algebra, $Y$ is substituted by
${\cal Y}$ and gradation $s$ is changed appropriately in
(\ref{barGrel}) and (\ref{Grel}).
Important difference, however, is that the chiral algebra
is typically not sufficient to reduce all correlators to those
of primaries alone. Unless some additional restrictions
are imposed, the input from concrete conformal model should
include, say, all triple vertices of the form
$\langle(W_{-1}^k V_\alpha)\ V_1 V_2\rangle$ in the case
of the $W_3$ algebra.
Therefore the analogue
(\ref{barGGvir}) can not be even written down in this form
and both vertices $\bar\Gamma$ and $\Gamma$ should be
evaluated separately.

\section{Examples of $3$-point functions
\label{3PF}}

\subsection{$\bar\Gamma$-type vertices}

\subsubsection{Virasoro
generators, eq.(\ref{barGrel})}

For primary $V_1$
and for $s=2$
we get from (\ref{barGrel}):
\be
\langle \underline{L_{-n} V_{\check\alpha}}\ |\ V_1(1)V_2(0) \rangle\ =
(n+1)\Delta_1\langle V_{\check\alpha}\ |\ V_1(1)V_2(0) \rangle\
+\langle V_{\check\alpha}\ |\ \underline{(L_{-1}V_1)(1)}\ V_2(0) \rangle\
+\langle V_{\check\alpha}\ |\ V_1(1)\ \underline{(L_nV_2)(0)} \rangle\
\label{barGrelV}
\ee
To make formulas more readable we underlined the entries,
where the Virasoro operators act on the vertex operators.

If $V_1$ is not a primary, one adds more terms at the r.h.s.:
\be
\langle \underline{L_{-n} V_{\check\alpha}}\ |\ V_1(1)V_2(0) \rangle\ =
\sum_{k>0} \frac{(n+1)!}{(k+1)!(n-k)!}
\langle V_{\check\alpha}\ |\ \underline{(L_kV_1)(1)}\ V_2(0) \rangle\
+ \\
+ (n+1)\Delta_1
\langle V_{\check\alpha}\ |\ V_1(1)V_2(0) \rangle\
+\langle V_{\check\alpha}\ |\ \underline{(L_{-1}V_1)(1)}\ V_2(0) \rangle\
+\ \langle V_{\check\alpha}\ |\ V_1(1)\ \underline{(L_nV_2)(0)} \rangle\
\label{barGrelVfull}
\ee
which, however, vanish at $n=0$.

If $V_2$ is a primary, then
for all $n>0$ one has $L_{n}V_2=0$ and then it follows from
(\ref{barGrelV}) that
\be
\langle V_{\check\alpha}\ |\ \underline{(L_{-1}V_1)(1)}\ V_2(0)\rangle \ =
\Big(\Delta_{\check\alpha} - \Delta_1-\Delta_2\Big)
\langle V_{\check\alpha}\ |\ V_1(1) V_2(0)\rangle
\label{L-1V1}
\ee
and
\be
\boxed{
\langle \underline{L_{-n} V_{\check\alpha}}\ |\ V_1(1)V_2(0) \rangle\ =
\Big(\Delta_{\check\alpha}  + n\Delta_1  - \Delta_2\Big)
\langle V_{\check\alpha}\ |\ V_1(1)V_2(0) \rangle,} \ \ \ \ n>0
\label{barvirins}
\ee
where we also used that $L_0 V_{\check\alpha}
= \Delta_{\check\alpha} V_{\check\alpha}$ and
$L_0 V_2 = \Delta_2 V_2$.
Note that (\ref{L-1V1}) remains valid when $V_1$ is
not obligatory a primary, while this is {\it not} true
for (\ref{barvirins}).
Note also that there were no restrictions on $V_{\check\alpha}$
in (\ref{barvirins}), in particular, $V_{\check\alpha}$
does not need to be a primary. If $V_{\check\alpha} = L_{-Y}V_\alpha$,
then
\be
\Delta_{\check\alpha} = \Delta_\alpha + |Y|
\label{dimY}
\ee
and one obtains from (\ref{barvirins}) for
$Y = \{k_1\geq k_2\geq\ldots$ and
$L_Y = \ldots L_{-k_2} \ldots L_{-k_1}$
\be
\boxed{
\langle \underline{L_{-Y} V_{\alpha}}\ |\ V_1(1)V_2(0) \rangle\ =
\langle V_{\alpha}\ |\ V_1(1)V_2(0) \rangle\
\prod_i \Big(\Delta_\alpha  + k_i\Delta_1  - \Delta_2 +
\sum_{j< i} k_j\Big),
}
\label{viriter}
\ee
which has been recently used in the check
of the $U(2)$ AGT relation in \cite{MMMagt}.

\subsubsection{$W^{(3)}$ generators, eq.(\ref{barGrel})}

Similarly, for the $W^{(3)}$-generator with $s=3$ and primaries $V_1$ and $V_2$
eq.(\ref{barGrel}) implies
\be
\langle \underline{W_{-n} V_{\check\alpha}}\ |\ V_1(1)V_2(0) \rangle\ =
\frac{(n+2)(n+1)}{2}w_1
\langle V_{\check\alpha}\ |\ V_1(1)V_2(0) \rangle\
+(n+2) \langle V_{\check\alpha}\ |\
\underline{(W_{-1}V_1)(1)}\ V_2(0) \rangle\
+\\
+\ \langle V_{\check\alpha}\ |\ \underline{(W_{-2}V_1)(1)}\ V_2(0) \rangle\
+\langle V_{\check\alpha}\ |\ V_1(1)\ \underline{(W_nV_2)(0)} \rangle\
\label{AAAbar}
\ee
This time one can exclude $\langle W_{-2}V_1\rangle$,
\be
\langle V_{\check\alpha}|\underline{(W_{-2}V_1)(1)}\ V_2(0) \rangle\ =
\Big(\hat w_{\check\alpha}-w_1-w_2\Big)
\langle V_{\check\alpha}|V_1(1)\ V_2(0) \rangle\ -
2\langle V_{\check\alpha}|\underline{(W_{-1}V_1)(1)}\ V_2(0) \rangle,
\label{AABbar}
\ee
from this system:
\be
\boxed{
\langle \underline{W_{-n}V_{\check\alpha}} | V_1(1)\ V_2(0) \rangle \
=
\left(\hat w_{\check\alpha}+ \frac{n(n+3)}{2}w_1-w_2\right)
\langle V_{\check\alpha} | V_1(1)\ V_2(0) \rangle\  +
n\langle V_{\check\alpha}  | \underline{(W_{-1}V_1)(1)}\ V_2(0)\rangle,}
\ \ n>0
\label{barWcor1}
\ee
One can also express the r.h.s. through
$W_{-1}V_{2}$ instead of $W_{-1}V_1$:
\be
\langle \underline{W_{-n}V_{\check\alpha}} | V_1(1)\ V_2(0) \rangle \
= (n+1)\left(\hat w_{\check\alpha}+ \frac{n}{2}w_1-w_2\right)
\langle V_{\check\alpha} | V_1(1)\ V_2(0) \rangle\  +
n\langle V_{\check\alpha}  | V_1(1)\ \underline{(W_{-1}V_2)(0)}
\rangle,  \ \ n>0
\label{barWcor2}
\ee
Here we made use of relation
\be\label{51}
\langle \underline{W_{-1}V_{\check\alpha}} | V_1(1)\ V_2(0) \rangle\
= \Big( \hat w_{\check\alpha}-w_1-w_2  \Big)
\langle V_{\check\alpha} | V_1(1)\ V_2(0) \rangle\
+ \langle V_{\check\alpha} | V_1(1)\ \underline{(W_{-1}V_2)(0)} \rangle\
\ee
The difference with the Virasoro case is that there are now
two structures at the r.h.s.

In fact, things are even more complicated.
Indeed, eqs.(\ref{barWcor1}) and (\ref{barWcor2}) are
written in a somewhat symbolical form which has a direct
meaning only for the primary $V_{\alpha}$.
Otherwise, what is denoted by
$\boxed{\hat w_{\check\alpha}V_{\check\alpha} \equiv W_0V_{\check\alpha}}$
is not proportional to $V_{\check\alpha}$: this is why we
put a hat over $w_{\check\alpha}$.
For example, already for the very first
descendants $L_{-1}V_\alpha$ and $W_{-1}V_\alpha$
\be
\hat w_{\alpha,L_{-1}}L_{-1}V_\alpha \equiv
W_0 (L_{-1}V_\alpha) = w_\alpha (L_{-1}V_\alpha)
+2(W_{-1}V_\alpha), \\
\hat w_{\alpha,W_{-1}}W_{-1}V_\alpha \equiv
W_0 (W_{-1}V_\alpha) = w_\alpha (W_{-1}V_\alpha) +
\frac{9D}{2} (L_{-1}V_\alpha)
\ee
are combinations of two different descendants (for the definition of
$D$ see (\ref{Ddef}) and (\ref{Ddef1}) below).
This makes it more difficult to write down for (\ref{barWcor1})
a unified iteration formula like (\ref{viriter}).

If $V_1$ and/or $V_2$ are not primaries, then additional terms
should be kept in (\ref{AAAbar}). We give just one example of
a full formula, where neither $V_{\check\alpha}$, nor $V_1$,
nor $V_2$ are assumed to be primaries:
\be
\langle \underline{W_{-1}V_{\check\alpha}} | V_1(1)\ V_2(0) \rangle \
= \ \langle \underline{W_{0}V_{\check\alpha}} | V_1(1)\ V_2(0) \rangle \
+ 2\langle V_{\check\alpha} | \underline{(W_{0}V_1)(1)}\ V_2(0) \rangle \
-\ \langle V_{\check\alpha} |
V_1(1)\ \underline{(W_{0}V_2)(0)} \rangle \ + \\
+ \ \langle V_{\check\alpha} | \underline{(W_{-1}V_1)(1)}\ V_2(0) \rangle \
+ \ \langle V_{\check\alpha} | \underline{(W_{1}V_1)(1)}\ V_2(0) \rangle \
+ \ \langle V_{\check\alpha} | V_1(1)\ \underline{(W_{1}V_2)(0)} \rangle
\label{barWcor1np}
\ee
For three primaries the three terms in the first line
combine into $w_\alpha+2w_1-w_2$ and the two last terms in the
second line disappear, thus reproducing (\ref{barWcor1})
with $n=1$.

\subsection{$\Gamma$-type vertices}

\subsubsection{Virasoro
generators, eq.(\ref{Grel})}

For primaries $V_3$ and $V_4$ and for $s=2$
one gets from (\ref{Grel})
\be
\langle \underline{(L_{-n}V_{\check\alpha})(0)}\ V_3(1)V_4(\infty) \rangle\ =
(n-1)\Delta_3\langle V_{\hat\alpha}(0)\ V_3(1)V_4(\infty)\rangle\
- \langle V_{\hat\alpha}(0)\ \underline{(L_{-1}V_3)(1)}\ V_4(\infty)\rangle\
+\\
+\ \langle V_{\hat\alpha}(0)\ V_3(1)\ \underline{L_{-n}V_4(\infty)}\rangle\
\label{Grel2L}
\ee
Taking $n> 0$ and excluding
\be
\langle V_{\check\alpha}(0)\ \underline{(L_{-1}V_3)(1)}\ V_4(\infty)\rangle =
-\Big(\Delta_{\check\alpha} +\Delta_3-\Delta_4\Big)
\langle V_{\check\alpha})(0)\ V_3(1)\ V_4(\infty)\rangle
\label{L-1V3}
\ee
from the resulting system, one obtains
\be
\boxed{
\langle \underline{(L_{-n} V_{\check\alpha})(0)}
\ V_3(1)V_4(\infty) \rangle\ =
\Big(\Delta_{\check\alpha}  + n\Delta_3  - \Delta_4\Big)
\langle V_{\check\alpha}(0)\  V_3(1)V_4(\infty) \rangle,} \ \ \ \ n>0
\label{virins}
\ee
i.e. exactly the same relation as (\ref{barvirins}).
Therefore, one also has
\be
\boxed{
\langle \underline{(L_{-Y} V_{\alpha})(0)}\ V_3(1)V_4(\infty) \rangle\ =
\langle V_{\alpha}(0)V_3(1)V_4(\infty) \rangle\
\prod_i \Big(\Delta_\alpha  + k_i\Delta_3  - \Delta_4 +
\sum_{j< i} k_j\Big),} \ \ \ \ n>0
\ee
and this validates the relation (\ref{barGGvir})
for the Virasoro chiral algebra.
Moreover, it follows from (\ref{Grel2L}) with $n<0$
and (\ref{L-1V3}) that
\be
\langle V_{\hat\alpha}(0)\ V_3(1)\ \underline{L_{-n}V_4(\infty)}\rangle\
= \Big(\Delta_4+n\Delta_3 -\Delta_{\check\alpha}\Big)
\langle V_{\hat\alpha}(0)\ V_3(1)\ V_4(\infty)\rangle,
\ \ \ \ n>0,
\ee
Note that signs in the similar equations
(\ref{L-1V1}) and (\ref{L-1V3}) are different.
Like (\ref{L-1V1}), eq.(\ref{L-1V3}) (but not
(\ref{virins})!) remains true when $V_1$ is any
descendant, not only a primary.

\subsubsection{$W^{(3)}$ generators, eq.(\ref{Grel})}

Similarly, for $W^{(3)}$-generator and primary $V_3$ and $V_4$,
eq.(\ref{Grel}) with $s=3$ implies that
\be
\langle \underline{(W_{-n}V_{\check\alpha})(0)}\
V_3(1)V_4(\infty) \rangle\ =
-\frac{(n-2)(n-1)}{2}\,w_3\,
\langle V_{\check\alpha}(0)\ V_3(1)V_4(\infty) \rangle\ +\\
+ (n-2)\langle V_{\check\alpha}(0)\
\underline{(W_{-1}V_3)(1)}\ V_4(\infty) \rangle\
-\langle V_{\check\alpha}(0)\ \underline{(W_{-2}V_3)(1)}\
V_4(\infty) \rangle\
- \ \langle V_{\check\alpha}(0)\ V_3(1)\
\underline{(W_nV_4)(\infty)}\rangle
\label{Grel2W}
\ee
Excluding  $\langle W_{-2}V_3\rangle$,
\be
\langle V_{\check\alpha}(0)\ \underline{(W_{-2}V_3)(1)}\
V_4(\infty) \rangle\ =
-\Big(\hat w_{\check\alpha}+w_3+w_4\Big)
\langle V_{\check\alpha}(0)\ V_3(1)V_4(\infty) \rangle\
- 2\langle V_{\check\alpha}(0)\ \underline{(W_{-1}V_3)(1)}\
V_4(\infty) \rangle,
\label{AAB}
\ee
from this system with $n\geq 0$, one obtains
\fr{
\langle \underline{(W_{-n}V_{\check\alpha})(0)}
\ V_3(1)\ V_4(\infty) \rangle \
=
 \left(\hat w_{\check\alpha} -\frac{n(n-3)}{2}w_3+w_4\right)
\langle V_{\check\alpha}(0)\ V_3(1)\ V_4(\infty) \rangle +\\
+ n\langle V_{\check\alpha}(0)\ \underline{(W_{-1}V_3)(1)}
\ V_4(\infty)\rangle,
\ \ \ \ n>0
\label{Wcor3}}
Again one can express the r.h.s. through
$W_{-1}V_{4}$ instead of $W_{-1}V_3$:
\be
\langle \underline{(W_{-n}V_{\check\alpha})(0)}\
V_3(1)\ V_4(\infty) \rangle \
(n+1)\left(\hat w_{\check\alpha} -\frac{n}{2}w_3+w_4\right)
\langle V_{\check\alpha}(0)\ V_3(1) V_4(\infty) \rangle \ -\\ -
n\langle V_{\check\alpha}(0)\ V_3(1)\
\underline{(W_{-1}V_4)(\infty)}\rangle,
\ \ \ \ n>0,
\label{Wcor4}
\ee
this time with the help of
\be
\langle V_{\check\alpha}(0)\ \underline{(W_{-1}V_3)(1)}
\ V_4(\infty) \rangle\ =
\Big(\hat w_{\check\alpha} - 2w_3 +w_4\Big)
\langle V_{\check\alpha}(0)\ V_3(1) V_4(\infty) \rangle \ -
\langle V_{\check\alpha}(0)\ V_3(1)\
\underline{(W_{-1}V_4)(\infty)} \rangle
\ee
Comparison with (\ref{barWcor1}) and (\ref{barWcor2})
shows that there is no clear substitute of the relation
(\ref{barGGvir}), which holds in the simple form
only for the Virasoro descendants.
Note also that if one uses (\ref{Grel2W}) to express
$\langle W_{-n}V_4\rangle$ for $n>0$, one obtains
\be
\langle V_{\check\alpha}(0)\ V_3(1)\
\underline{(W_{-n}V_4)(\infty)} \rangle\ =
\left(w_4-\frac{n(n+3)}{2}w_3+\hat w_{\check\alpha}\right)
\langle V_{\check\alpha}(0)\ V_3(1)V_4(\infty) \rangle\ -\\
- n\langle V_{\check\alpha}(0)\
\underline{(W_{-1}V_3)(1)}\ V_4(\infty) \rangle,
\ \ \ n>0
\ee
which is very close, though still different from (\ref{barWcor1}).

The counterpart of eq.(\ref{barWcor1np})
for three generic (not obligatory primary) operators is now
\be
\langle \underline{(W_{-1}V_{\check\alpha})(0)}
\ V_3(1)\ V_4(0) \rangle \
= \ \langle \underline{(W_{0}V_{\check\alpha})(0)}
\ V_3(1)\ V_4(0) \rangle \
+ \langle V_{\check\alpha}(0)\ \underline{(W_{0}V_3)(1)}
\ V_4(0) \rangle \ +\\
+ \langle V_{\check\alpha}(0)\ V_3(1)\
\underline{(W_{0}V_4)}(0) \rangle \
+  \langle V_{\check\alpha}(0)\ \underline{(W_{-1}V_3)(1)}
\ V_4(0) \rangle \
- \ \langle V_{\check\alpha}(0)\ V_3(1)\
\underline{(W_{1}V_4)(0)} \rangle
\label{Wcor3np}
\ee
For three primaries, the three terms in the first line
combine into $w_\alpha+w_3+w_4$ and the very last term in the
second line disappears, thus reproducing (\ref{Wcor3})
with $n=1$. Note that, in variance with (\ref{barWcor1np}),
there is no term
$\langle V_{\check\alpha}(0)\ (W_{1}V_3)(1)\ V_4(0) \rangle$
in the second line of (\ref{Wcor3np}): this is specifics of our
restriction to $n=1$ in this example.

\section{Example of free field calculation}

Now we repeat the somewhat abstract calculations of the
previous sections in the model of free fields,
where all operators can be explicitly defined and their
correlators can be explicitly calculated.
This would help to illustrate formulas and also to
check that they are correct.
We also restrict consideration to particular correlators with
$W$-operator found above.
Since we did not go beyond the $W^{(3)}$ algebra,
we need to consider only two free fields at most.
Our final restriction is to the case of the central charge
$c=2$, i.e. $\vec Q=0$: formulas are somewhat lengthy
even in this case and overloading them further would
jeopardize their use for illustrative purposes.
An arbitrary $Q$ appears only in our illustrations of
correlators of Virasoro-descendants with a single free field.
All generalizations are absolutely straightforward.

\subsection{Decomposition rule (\ref{4p23pinf})}

Let us begin from eq.(\ref{4p23pinf}).
The starting statement is that
for $\vec\alpha_1+\vec\alpha_2+\vec\alpha_3+\vec\alpha_4 = 0$
the correlator
\be
x^{-\vec\alpha_1\vec\alpha_2}\langle \no e^{\vec\alpha_1\vec\phi(x)}\no \
\no e^{\vec\alpha_2\vec\phi(0)}\no \ \no e^{\vec\alpha_3\vec\phi(1)}\no \
\no e^{\vec\alpha_4\vec\phi(\infty)}\no \rangle\ \sim
(1-x)^{\vec\alpha_1\vec\alpha_3}  =
1 - (\vec\alpha_1\vec\alpha_3)x + \frac{(\vec\alpha_1\vec\alpha_3)
(\vec\alpha_1\vec\alpha_3 - 1)}{2}\,x^2 + \ldots
\label{frefco}
\ee
can also be decomposed with the help of (\ref{ope}) as
\be
\langle \no e^{\vec\alpha_1\vec\phi(x)+\vec\alpha_2\vec\phi(0)}\no \
\no e^{\vec\alpha_3\vec\phi(1)}\no \
\no e^{\vec\alpha_4\vec\phi(\infty)}\no \rangle\ =
\langle \no e^{(\vec\alpha_1+\vec\alpha_2)\vec\phi(0)}\no \
\no e^{\vec\alpha_3\vec\phi(1)}\no \
\no e^{\vec\alpha_4\vec\phi(\infty)}\no \rangle\ +\\\ +
x \langle \no (\vec\alpha_1\p\vec\phi)
e^{(\vec\alpha_1+\vec\alpha_2)\vec\phi}(0)\no \
\no e^{\vec\alpha_3\vec\phi(1)}\no \
\no e^{\vec\alpha_4\vec\phi(\infty)}\no \rangle\ + \\ +
\frac{x^2}{2} \langle \no \Big((\vec\alpha_1\p\vec\phi)^2 +
\vec\alpha_1\p^2\vec\phi\Big)
e^{(\vec\alpha_1+\vec\alpha_2)\vec\phi}(0)\no \
\no e^{\vec\alpha_3\vec\phi(1)}\no \
\no e^{\vec\alpha_4\vec\phi(\infty)}\no \rangle\ +
\ldots
\label{expoexpan}
\ee
and further expanded in the Virasoro/$W$-algebra-related basis
\be
\langle \no e^{\vec\alpha_1\vec\phi(x)+\vec\alpha_2\vec\phi(0)}\no \
\no e^{\vec\alpha_3\vec\phi(1)}\no \
\no e^{\vec\alpha_4\vec\phi(\infty)}\no \rangle\ =
\underline{C_{\alpha_1\alpha_2}^\alpha
\langle \no e^{(\vec\alpha_1+\vec\alpha_2)\vec\phi(0)}\no \
\no e^{\vec\alpha_3\vec\phi(1)}\no \
\no e^{\vec\alpha_1\vec\phi(\infty)}\no \rangle}\ + \\ +
x\left\{\begin{array}{c}
\underline{C_{\alpha_1\alpha_2}^{\alpha\!,\,L_{-1}}
\langle \left(L_{-1}\no e^{(\vec\alpha_1+\vec\alpha_2)\vec\phi(0)}\no \right)\
\no e^{\vec\alpha_3\vec\phi(1)}\no \
\no e^{\vec\alpha_4\vec\phi(\infty)}\no \rangle}\ +  \\ +\
C_{\alpha_1\alpha_2}^{\alpha\!,\,W_{-1}}
\langle \left(W_{-1}\no e^{(\vec\alpha_1+\vec\alpha_2)\vec\phi(0)}\no \right)
\no e^{\vec\alpha_3\vec\phi(1)}\no \
\no e^{\vec\alpha_4\vec\phi(\infty)}\no \rangle\ +
\end{array}\right.
\\ +
x^2\left\{\begin{array}{c}
\underline{C_{\alpha_1\alpha_2}^{\alpha\!,\,L_{-2}}
\langle \left(L_{-2}\no e^{(\vec\alpha_1+\vec\alpha_2)\vec\phi(0)}\no \right)\
\no e^{\vec\alpha_3\vec\phi(1)}\no \
\no e^{\vec\alpha_4\vec\phi(\infty)}\no \rangle}\ +  \\ +\
\underline{C_{\alpha_1\alpha_2}^{\alpha\!,\,L^2_{-1}}
\langle \left(L_{-1}^2\no e^{(\vec\alpha_1+\vec\alpha_2)\vec\phi(0)}\no \right)
\no e^{\vec\alpha_3\vec\phi(1)}\no \
\no e^{\vec\alpha_4\vec\phi(\infty)}\no \rangle}\ + \\ +\
C_{\alpha_1\alpha_2}^{\alpha\!,\,L_{-1}W_{-1}}
\langle \left(L_{-1}W_{-1}\no e^{(\vec\alpha_1+\vec\alpha_2)\vec\phi(0)}\no \right)
\no e^{\vec\alpha_3\vec\phi(1)}\no \
\no e^{\vec\alpha_4\vec\phi(\infty)}\no \rangle\ + \\ +\
C_{\alpha_1\alpha_2}^{\alpha\!,\,W_{-2}}
\langle \left(W_{-2}\no e^{(\vec\alpha_1+\vec\alpha_2)\vec\phi(0)}\no \right)
\no e^{\vec\alpha_3\vec\phi(1)}\no \
\no e^{\vec\alpha_4\vec\phi(\infty)}\no \rangle\ + \\ +\
C_{\alpha_1\alpha_2}^{\alpha\!,\,W^2_{-1}}
\langle \left(W_{-1}^2\no e^{(\vec\alpha_1+\vec\alpha_2)\vec\phi(0)}\no \right)
\no e^{\vec\alpha_3\vec\phi(1)}\no \
\no e^{\vec\alpha_4\vec\phi(\infty)}\no \rangle\ +
\end{array}\right.
\\ +
\ldots
\label{expoexpanLW}
\ee
Our goal in this section is to show in detail how these two
expansions match each other. This is important because free
fields provide a rare example when not only the second
expansion (\ref{expoexpanLW}) can be explicitly written down,
but also the first one, (\ref{frefco}) is known, and one can
see how sophisticated formula (\ref{expoexpanLW}) manages
to reproduce the simple answer (\ref{frefco}), order by order
in the $x$ expansion.

\subsubsection{The first-level term in the case of a single free field
with $c=1$}

It is very simple to deal with the first-level term in the
case when there is just a single field $\phi$: then
only the underlined terms contribute in (\ref{expoexpanLW}),
and, since
\be
(\alpha_1\p\phi)e^{(\alpha_1+\alpha_2)\phi}(0) =
\frac{\alpha_1}{\alpha_1+\alpha_2}
\left(\p\, e^{(\alpha_1+\alpha_2)\phi}\right)(0)
\ee
the structure constant in this case is simply
\be
C^{\,(\alpha\!,L_{-1})}_{\alpha_1\alpha_2}= \frac{\alpha_1}{\alpha_1+\alpha_2}
\,\delta_{\alpha,\alpha_1+\alpha_2},
\label{Clev11f}
\ee
Here there is a specific free field theory selection rule
\be
\alpha=\alpha_1+\alpha_2
\ee
We now need to calculate the triple vertex
$\gamma_{\alpha\alpha_3\alpha_4}(L_{-1})$.
From the general formula for a triple correlator
of primaries,
\be
\langle V(z) V_3(z_3) V_4(z_4)\rangle\ \sim
\frac{1}{(z-z_{3})^{\Delta+\Delta_3-\Delta_4}
(z-z_{4})^{\Delta+\Delta_4-\Delta_3} z_{34}^{\Delta_3+\Delta_4-\Delta}}
\ee
it follows by taking $z$-derivative that
\be
\langle (\p V)(z)\ V_3(z_3) V_4(z_4)\rangle\ =\
-\langle V(z) V_3(z_3) V_4(z_4)\rangle\left(
\frac{\Delta+\Delta_3-\Delta_4}{z-z_{3}} +
\frac{\Delta+\Delta_4-\Delta_3}{z-z_{4}}\right) \longrightarrow\\
\ \stackrel{z_4\rightarrow\infty}{\longrightarrow} \
-\frac{\Delta+\Delta_3-\Delta_4}{z-z_{3}}
\langle V(z) V_3(z_3) V_4(z_4)\rangle\ \
\ \stackrel{z=0,\ z_3=1}{\longrightarrow} \
\Big(\Delta+\Delta_3-\Delta_4\Big)
\langle V(z) V_3(z_3) V_4(z_4)\rangle,
\ee
i.e., in nice accordance with (\ref{barvirins}),
\be
\gamma_{\alpha\alpha_3\alpha_4}(L_{-1}) = \Delta+\Delta_3-\Delta_4
\ee

Thus matching of the two sides of (\ref{4p23pinf})
in this very particular case implies that
\be
-\alpha_1\alpha_3 =
\sum_\alpha C^{\,(\alpha\!,L_{-1})}_{\alpha_1\alpha_2}
\gamma_{\alpha\alpha_3\alpha_4}(L_{-1})
= \frac{\alpha_1\Big(\Delta_{\alpha_1+\alpha_2}+\Delta_3-\Delta_4\Big)}
{\alpha_1+\alpha_2}
\label{primer1}
\ee
One now substitutes $\Delta_\alpha = \frac{\alpha^2}{2}$
and obtains
\be
\Delta_{\alpha_1+\alpha_2}+\Delta_3-\Delta_4
= \frac{(\alpha_1+\alpha_2)^2+\alpha_3^2 -\alpha_4^2}{2}
= \frac{(\alpha_3+\alpha_4)^2+\alpha_3^2 -\alpha_4^2}{2}
= \alpha_3(\alpha_3+\alpha_4)
\label{dimcomp1}
\ee
where at the first step the free field theory selection
rule $\alpha_1+\alpha_2+\alpha_3+\alpha_4=0$ is used.
Using this rule once again, one obtains that (\ref{primer1})
is indeed true.

This is an example of {\it explicit} calculation,
which checks general formulas for triple vertices.
Note that during this check the details of the model were important,
especially the selection (charge conservation) rules.
This is because we explicitly calculated all the correlators,
what is possible only in a given model with all its
specific properties. The relations which we want to illustrate,
like (\ref{4p23pinf}) and (\ref{barvirins}) are model-independent,
therefore our consideration can only illustrate and validate
them, but not prove.
The proof (derivation) was given in the previous section,
it is instead a little more abstract.

Now we proceed to generalizations of this example.
They go in three directions: to $Q\neq 0$,
to higher descendant level (size of the Young diagram)
and to several free fields and, hence, extended chiral algebras.
We make only one step in each direction.

\subsubsection{The first-level term in the case of a single free field
with $c\neq 1$}

This generalization is trivial. Nothing changes except for
the formula for dimensions: $\Delta_\alpha = \frac{(\alpha-Q)^2-Q^2}{2}$
and the selection rule $\alpha+\alpha_3+\alpha_4=2Q$.
Note that the other selection rule, $\alpha=\alpha_1+\alpha_2$
remains intact.
Then instead of (\ref{dimcomp1}) one gets
\be
\Delta_{\alpha}+\Delta_3-\Delta_4 =
\Delta_{2Q-\alpha_3-\alpha_4}+\Delta_3-\Delta_4 =
\frac{(\alpha_3+\alpha_4-Q)^2+(\alpha_3-Q)^2-(\alpha_4-Q)^2-Q^2}{2}=\\
= \alpha_3(\alpha_3+\alpha_4-2Q) = -\alpha\alpha_3
= -\alpha_3(\alpha_1+\alpha_2),
\ee
what after substitution into the r.h.s. of (\ref{primer1})
makes it into identity.

\subsubsection{The second-level term in the case of a single free field
with $c=1$}

At this level, there are two independent operators
$\ \no (\p\phi)^2e^{\alpha\phi}\no \ $ and $\ \no \p^2\phi e^{\alpha\phi}\no  $
We want, however, to use another basis for linear decomposition
of level-two operators: that formed by $L_{-2}\no e^{\alpha\phi}\no \ $
and $L_{-1}^2\no e^{\alpha\phi}\no \ $
To use this basis, one needs to find out how $L_{-1}^2$ acts on
$\ \no e^{\alpha\phi}\no \ $ This follows from the operator product expansion
\be
T(x) \Big(L_{-1}\no e^{\alpha\phi}\no \!\!\Big)(0)\
=\frac{1}{2}\no (\partial\phi)^2(x)\no \ \no \alpha\partial\phi e^{\alpha\phi}(0)\no \
= \\
=\frac{\alpha^2}{2x^3} \no e^{\alpha\phi}(0)\no \ +
\frac{\alpha^3+2\alpha}{2x^2} \no \partial\phi e^{\alpha\phi}(0)\no \ +\
\frac{\alpha}{x} \no (\partial^2\phi+\alpha(\partial\phi)^2)e^{\alpha\phi}(0)\no
\ +\ldots
\ee
Picking up terms with various powers of $x$, one obtains that
\be
L_1L_{-1}\no e^{\alpha\phi}\no \ = \frac{\alpha^2}{2}\no e^{\alpha\phi}\no \
= \Delta_\alpha \no e^{\alpha\phi}\no  \\
L_0L_{-1}\no e^{\alpha\phi}\no \ = \frac{\alpha^3+2\alpha}{2}
\no \partial\phi e^{\alpha\phi}(0)\no \ = (\Delta_\alpha+1)
L_{-1}\no e^{\alpha\phi}\no \\
L_{-1}^2\no e^{\alpha\phi}\no \ =
\no \Big(\alpha\partial^2\phi+(\alpha\partial\phi)^2\Big)e^{\alpha\phi}\no \
= \p^2\no e^{\alpha\phi}\no
\ee
i.e. $L_1V_\alpha$ has dimension $\Delta_\alpha+1$ and
$L_{-1}^2$ acts on $V_\alpha$ as a second derivative.
Both conclusions are actually true not only for primaries,
and not only for the free field model.
We also know from (\ref{Tonexp}) that
\be
L_{-2}\no e^{\alpha\phi}\no \ \stackrel{(\ref{Tonexp})}{=}\
\no \left(\frac{1}{2}(\partial\phi)^2+\alpha\partial^2\phi\right)
e^{\alpha\phi}\no
\ee
Coming back to (\ref{expoexpan}), at the second level one
represents
\be
\frac{1}{2}\Big(\alpha_1^2(\partial\phi)^2+\alpha_1\partial^2\phi
\Big) =
C_{\alpha_1,\alpha_2}^{\alpha,L_{-2}}\,
\left(\frac{1}{2}(\partial\phi)^2+\alpha\partial^2\phi\right)
\ +\
C_{\alpha_1,\alpha_2}^{\alpha,L_{-1}^2}\,
\Big(\alpha\partial^2\phi+\alpha^2(\partial\phi)^2\Big)
\ee
which means that
\be
C_{\alpha_1,\alpha_2}^{\alpha,L_{-2}}=\frac{\alpha_{1}(\alpha_1-\alpha)}{1-2\alpha^2}
\\
C_{\alpha_1,\alpha_2}^{\alpha,L_{-1}^2}=
\frac{\alpha_{1}(2\alpha_1\alpha-1)}{4\alpha^3-2\alpha}
\label{Clev21f}
\ee
As the next step, one needs the corresponding triple vertices $\gamma$, which are
\be
\langle \left(L_{-2}\no e^{(\vec\alpha_1+\vec\alpha_2)\vec\phi(0)}\no \right)\
\no e^{\vec\alpha_3\vec\phi(1)}\no \
\no e^{\vec\alpha_4\vec\phi(\infty)}\no \rangle\ =
\left(\frac{\alpha_3^2}{2}-\alpha_3\alpha \right)\langle
\no e^{(\vec\alpha_1+\vec\alpha_2)\vec\phi(0)}\no \
\no e^{\vec\alpha_3\vec\phi(1)}\no \
\no e^{\vec\alpha_4\vec\phi(\infty)}\no \rangle \\
\langle \left(L_{-1}^2\no e^{(\vec\alpha_1+\vec\alpha_2)\vec\phi(0)}\no \right)
\no e^{\vec\alpha_3\vec\phi(1)}\no \
\no e^{\vec\alpha_4\vec\phi(\infty)}\no \rangle\ =
\Big((\alpha_3\alpha)^2-\alpha_3\alpha\Big)\langle
\no e^{(\vec\alpha_1+\vec\alpha_2)\vec\phi(0)}\no
\no e^{\vec\alpha_3\vec\phi(1)}\no \
\no e^{\vec\alpha_4\vec\phi(\infty)}\no \rangle\
\ee
Now we have everything to check, using (\ref{expoexpanLW}), the matching condition
between (\ref{frefco}) and (\ref{expoexpan}) at level two:
\be
\frac{(\alpha_1\alpha_3)
(\alpha_1\alpha_3 - 1)}{2} =
\Big(\frac{\alpha_{1}(\alpha_1-\alpha)}{1-2\alpha^2}\Big)
\Big(\frac{\alpha_3^2}{2}-\alpha_3\alpha\Big)+
\Big(\frac{\alpha_{1}(2\alpha_1\alpha-1)}{4\alpha^3-2\alpha}\Big)
\Big((\alpha_3\alpha)^2-\alpha_3\alpha\Big)
\ee
which is, indeed, correct provided $\alpha=\alpha_1+\alpha_2$.

\subsubsection{The  first-level term in the case of two free
fields with $c=2$}

In the case of several fields
$(\vec\alpha_1\p\vec\phi) e^{(\vec\alpha_1+\vec\alpha_2)\vec\phi}$
is no longer a derivative of the exponential primary field.
Already at the first level one needs to decompose it into
several descendant operators.
In the case of two fields, $r=2$, we denote
$\vec\alpha\vec\phi(z) = \alpha\phi_1(z)+\beta\phi_2(z)$.
Then
\be
T(x)\ \no e^{\vec\alpha\vec\phi(0)}\no \ =
\frac{1}{2}\no \Big( (\p\phi_1)^2(x) + (\p\phi_2)^2(x)\Big)\no
\ \no e^{(\alpha\phi_1+\beta\phi_2)(0)}\no \ =
\frac{\alpha^2+\beta^2}{2x^2}\ \no e^{(\alpha\phi_1+\beta\phi_2)(0)}\no \
+  \\ +
\frac{1}{x}\no \!(\alpha\p\phi_1 + \beta\p\phi_2)
e^{(\alpha\phi_1+\beta\phi_2)}(0)\no \ +
\no \left( \frac{1}{2}(\p\phi_1)^2 + \frac{1}{2}(\p\phi_2)^2 +
\alpha\p^2\phi_1 + \beta\p^2\phi_2\right)
e^{(\alpha\phi_1+\beta\phi_2)}(0)\no \ + O(x)
\label{T-exp}
\ee
At the same time, by definition of Virasoro operators $L_n$
acting on operators at point $z=0$,
\be
T(x)\ \no e^{\vec\alpha\vec\phi(0)}\no \ =
\sum_k \frac{1}{x^{k+2}}\ L_k\! \no e^{\vec\alpha\vec\phi}(0)\no
\label{Texpan}
\ee
thus, one obtains
\be
L_n\, \no e^{\vec\alpha\vec\phi}\no \ = 0, \ \ \ \ \ \ {\rm for}\ n>0,\\
L_0\, \no e^{\vec\alpha\vec\phi}\no \ = \Delta_{\alpha,\beta}
\no e^{\vec\alpha\vec\phi}\no \ =
\frac{\alpha^2+\beta^2}{2}
\no e^{\vec\alpha\vec\phi}\no  \\
L_{-1}\, \no e^{\vec\alpha\vec\phi}\no \ =\
\no \!(\alpha\p\phi_1 + \beta\p\phi_2)
e^{(\alpha\phi_1+\beta\phi_2)}\no  \\
L_{-2}\, \no e^{\vec\alpha\vec\phi}\no \ =
\no \!\left(\frac{1}{2} (\p\phi_1)^2 + \frac{1}{2}(\p\phi_2)^2 +
\alpha\p^2\phi_1 + \beta\p^2\phi_2\right)
e^{(\alpha\phi_1+\beta\phi_2)}\no
\label{L-exp}
\ee
However, already at the first level one does not obtain the
complete basis enough to the decompose arbitrary operator
$\no (A\p\phi_1+B\p\phi_2)e^{\vec\alpha\vec\phi}\no $, only the
particular direction $B/A=\beta/\alpha$ in this
two-dimensional space is spanned by $L_{-1}\no e^{\vec\alpha\vec\phi}\no $
The second vector in the basis, $W_{-1}\no e^{\vec\alpha\vec\phi}\no $
is produced by the action of the $W^{(3)}$ operator.
Since we consider only the case of two fields, there will
be no other $W$-operators and we denote it simply by $W(z)$.
For $W^{(2)}(z)$ we already have a special notation:
the stress tensor is nothing but $W^{(2)}(z)=T(z)$.

$W(z)$ is cubic in field derivatives $\p\phi(z)$,
and additional requirement is that the most singular term
in its operator product expansion with $T(z)$ is absent.
The stress tensor $T(z)$ is invariant under
$SO(2)$ rotations of $\vec\phi$, and this rotation freedom
should be fixed in order to define $W(z)$ unambiguously.
We require that this latter is symmetric under $\phi_2 \rightarrow -\phi_2$
and, therefore, antisymmetric under $\phi_1\rightarrow -\phi_1$.
This means that $W = (\p\phi_1)^3 + h \p\phi_1(\p\phi_2)^2$
with a single undefined parameter $h$.
The operator product expansion is
\be
T(z)W(0) = \frac{3+h}{z^4}\p\phi_1(0) + \ldots
\ee
and the additional requirement is that the most singular term with $z^{-4}$
is absent, or, to put it differently, that $W(z)$ is a primary of the Virasoro
algebra. This defines $h$ to be $h=-3$,
\be
W(z) = W^{(3)}(z) \equiv (\p\phi_1)^3 -3\p\phi_1(\p\phi_2)^2
= \p\phi_1\Big((\p\phi_1)^2 - 3(\p\phi_2)^2\Big)
\ee
and
\be
T(z)W(0) = \frac{3W(0)}{z^2} + \frac{\p W(0)}{z} + \ldots
\label{TWope}
\ee
so that the conformal dimension of $W(z)$ is 3.

A counterpart of (\ref{T-exp}) for the $W$-operator is
\be
W(x) \ \no e^{\vec\alpha\vec\phi(0)}\no \ =
\frac{\alpha(\alpha^2-3\beta^2)}{x^3}\, \no e^{\vec\alpha\vec\phi(0)}\no \
+ \frac{1}{x^2}\no \!\Big(3(\alpha^2-\beta^2)\p\phi_1
- 6\alpha\beta \p\phi_2\Big)
e^{(\alpha\phi_1+\beta\phi_2)}(0)\no  \ + \\ +
\frac{1}{x}
\ \no 3\Big(\alpha(\p\phi_1)^2 -2\beta \p\phi_1\p\phi_2 - \alpha(\p\phi_2)^2
+ (\alpha^2-\beta^2)\p^2\phi_1 - 2\alpha\beta \p^2\phi_2\Big)
e^{(\alpha\phi_1+\beta\phi_2)}\no  \ + O(1)
\label{Wop-exp}
\ee
Components (harmonics) of the $W$-operator are defined in direct
analogy with the Virasoro generators in (\ref{Texpan}):
\be
W(x)\ \no e^{\vec\alpha\vec\phi(0)}\no \ =
\sum_k \frac{1}{x^{k+3}}\ W_k\! \no e^{\vec\alpha\vec\phi}(0)\no
\label{Wexpan}
\ee
and, therefore,
\be
W_n\, \no e^{\vec\alpha\vec\phi}\no \ = 0, \ \ \ {\rm for}\ n>0,\\
W_0\, \no e^{\vec\alpha\vec\phi}\no \ = w_{\alpha,\beta}
\no e^{\vec\alpha\vec\phi}\no \ =
\alpha(\alpha^2-3\beta^2)
\no e^{\vec\alpha\vec\phi}\no  \\
W_{-1}\, \no e^{\vec\alpha\vec\phi}\no \ =\
\no \!3\Big((\alpha^2-\beta^2)\p\phi_1 - 2\alpha\beta \p\phi_2\Big)
e^{(\alpha\phi_1+\beta\phi_2)}\no \\
W_{-2}\, \no e^{\vec\alpha\vec\phi}\no \ =
\ \no 3\Big(\alpha(\p\phi_1)^2 -2\beta \p\phi_1\p\phi_2 - \alpha(\p\phi_2)^2
+ (\alpha^2-\beta^2)\p^2\phi_1 - 2\alpha\beta \p^2\phi_2\Big)
e^{(\alpha\phi_1+\beta\phi_2)}\no
\label{W-exp}
\ee
Combining the relevant lines in (\ref{L-exp}) and (\ref{W-exp}),
one obtains that at level one
\be
\no \p\phi_1 e^{(\alpha\phi_1+\beta\phi_2)}\no  \ =
\frac{1}{3(\alpha^2-\beta^2)}\Big(
6\alpha L_{-1} + W_{-1}\Big)e^{(\alpha\phi_1+\beta\phi_2)}\no \\
\no \p\phi_2 e^{(\alpha\phi_1+\beta\phi_2)}\no  \ =
\frac{1}{3\beta(\alpha^2-\beta^2)}\Big(
3(\alpha^2-\beta^2) L_{-1} - \alpha W_{-1}\Big)
e^{(\alpha\phi_1+\beta\phi_2)}\no
\ee
so that the operator which appears at the first level in
(\ref{expoexpan}) can be represented as
\be
\no (\vec\alpha_1\p\vec\phi) e^{(\alpha\phi_1+\beta\phi_2)}\no  \ =
\left(\frac{\Big(6\alpha\beta\alpha_1 + 3(\alpha^2-\beta^2)\beta_1\Big)}
{3\beta(3\alpha^2-\beta^2)}L_{-1}
+ \frac{\alpha_1\beta-\alpha\beta_1}{3\beta(3\alpha^2-\beta^2)}W_{-1}
\right)\no e^{(\alpha\phi_1+\beta\phi_2)}\no
\ee
Therefore, we calculated the first two structure
constants in (\ref{expoexpanLW}), in addition to the trivial one
$C_{\alpha_1\alpha_2}^\alpha = 1$:
\be
C_{\alpha_1\alpha_2}^{\alpha\!,\,L_{-1}} =
\frac{\Big(6\alpha\beta\alpha_1 + 3(\alpha^2-\beta^2)\beta_1\Big)}
{3\beta(3\alpha^2-\beta^2)},\\
C_{\alpha_1\alpha_2}^{\alpha\!,\,W_{-1}} =
\frac{\alpha_1\beta-\alpha\beta_1}{3\beta(3\alpha^2-\beta^2)}
\label{Clev1}
\ee
Since we also know the explicit expressions for
$L_{-1}\, \no e^{\vec\alpha\vec\phi}\no \ $ and
$W_{-1}\, \no e^{\vec\alpha\vec\phi}\no $, one can explicitly evaluate
the three-point functions
\be
\Gamma_{\vec\alpha\vec\alpha_3\vec\alpha_4}(L_{-1}) \equiv\
\langle \left(L_{-1}\! \no e^{\vec\alpha\vec\phi}(0)\no \right)\
\no e^{\vec\alpha_3\vec\phi(1)}\no \
\no e^{\vec\alpha_4\vec\phi(\infty)}\no \rangle\
\stackrel{(\ref{L-exp})}{=}\
\langle \no (\vec\alpha \p\vec\phi) e^{\vec\alpha\vec\phi}(0)\no \
\no e^{\vec\alpha_3\vec\phi(1)}\no \
\no e^{\vec\alpha_4\vec\phi(\infty)}\no \rangle\ = \\ =
- (\vec\alpha\vec\alpha_3)
\langle \no e^{\vec\alpha\vec\phi}(0)\no \
\no e^{\vec\alpha_3\vec\phi(1)}\no \
\no e^{\vec\alpha_4\vec\phi(\infty)}\no \rangle\
\ee
and
\be
\Gamma_{\vec\alpha\vec\alpha_3\vec\alpha_4}(W_{-1}) \equiv\
\langle \left(W_{-1}\! \no e^{\vec\alpha\vec\phi}(0)\no \right)\
\no e^{\vec\alpha_3\vec\phi(1)}\no \
\no e^{\vec\alpha_4\vec\phi(\infty)}\no \rangle\
\stackrel{(\ref{W-exp})}{=}\\
\langle \no \!\Big(3(\alpha^2-\beta^2)\p\phi_1 - 6\alpha\beta \p\phi_2\Big)
e^{\vec\alpha\vec\phi}(0)\no \
\no e^{\vec\alpha_3\vec\phi(1)}\no \
\no e^{\vec\alpha_4\vec\phi(\infty)}\no \rangle\ = \\
= -\Big( 3(\alpha^2-\beta^2)\alpha_3 - 6\alpha\beta\beta_3\Big)
\langle \no e^{\vec\alpha\vec\phi}(0)\no \
\no e^{\vec\alpha_3\vec\phi(1)}\no \
\no e^{\vec\alpha_4\vec\phi(\infty)}\no \rangle
\label{Wlev1}
\ee
Here we took into account that pairings with an operator at
infinity give rise to an extra factor of $(0-\infty)$
in denominator and, therefore, can be neglected.
Pairings with an operator at $z_3=1$ produce a factor of
$(z-z_3)^{-1} = -1$, responsible for the minus sign in the both
formulas. Thus,
\be
\gamma_{\vec\alpha\vec\alpha_3\vec\alpha_4}(L_{-1})
= -(\alpha\alpha_3+\beta\beta_3), \\
\gamma_{\vec\alpha\vec\alpha_3\vec\alpha_4}(W_{-1})
= 3\Big(-(\alpha^2-\beta^2)\alpha_3 + 2\alpha\beta\beta_3\Big)
\label{gammalev1}
\ee
These formulas are in accordance with (\ref{virins}) and
(\ref{Wcor3}) respectively, and this can serve as a check
of those general expressions.
Indeed, by (\ref{virins})
\be
\gamma_{\vec\alpha\vec\alpha_3\vec\alpha_4}(L_{-1})
\ \stackrel{(\ref{virins})}{=} \Delta_{\vec\alpha}
+ \Delta_{\vec\alpha_3} -\Delta_{\vec\alpha_4}
= \frac{(\vec\alpha_3+\vec\alpha_4)^2+\vec\alpha_3^2-\vec\alpha_4^2}{2}
= \vec\alpha_3(\vec\alpha_3+\vec\alpha_4) = -\vec\alpha\vec\alpha_3
\ee
where the selection rule $\vec\alpha+\vec\alpha_3+\vec\alpha_4=\vec 0$
was used twice.
Similarly, by (\ref{Wcor3})
\be\label{101}
\gamma_{\vec\alpha\vec\alpha_3\vec\alpha_4}(W_{-1})
\ \stackrel{(\ref{gammalev1})}{=}\
3\Big(-(\alpha^2-\beta^2)\alpha_3 + 2\alpha\beta\beta_3\Big)
\ \stackrel{(\ref{virins})}{=}\ w_{\vec\alpha} + w_{\vec\alpha_3}
+w_{\vec\alpha_4} +
3\Big( (\alpha_3^2-\beta_3^2)\alpha - 2\alpha_3\beta_3\beta\Big),
\ee
The last equality is, indeed, an identity
provided $\vec\alpha+\vec\alpha_3+\vec\alpha_4=0$.
The last bracket at the r.h.s. is a direct counterpart
of (\ref{Wlev1}) for another position of the $W_{-1}$ operator,
which can also be explicitly evaluated in the free field theory:
\be
\langle \no e^{\vec\alpha\vec\phi}(0)\no \ \left(W_{-1}\! \no e^{\vec\alpha_3\vec\phi(1)}\no \right)\
\no e^{\vec\alpha_4\vec\phi(\infty)}\no \rangle\
\stackrel{(\ref{W-exp})}{=}\\
\langle \no e^{\vec\alpha\vec\phi}(0)\no \
\no \!\Big(3(\alpha_3^2-\beta_3^2)\p\phi_1 - 6\alpha_3\beta_3 \p\phi_2\Big)
e^{\vec\alpha_3\vec\phi(1)}\no \
\no e^{\vec\alpha_4\vec\phi(\infty)}\no \rangle\ =  \\
= \Big( 3(\alpha_3^2-\beta_3^2)\alpha - 6\alpha_3\beta_3\beta\Big)
\langle \no e^{\vec\alpha\vec\phi}(0)\no \
\no e^{\vec\alpha_3\vec\phi(1)}\no \
\no e^{\vec\alpha_4\vec\phi(\infty)}\no \rangle
\label{Wlev13}
\ee
Similarly to (\ref{Wlev1}), only the pairings between operators at points
$z=0$ and $z_3=1$ are contributing. This time the derivative is taken
w.r.t. $z_3$, therefore, there is no overall minus sign.

Finally, one can combine (\ref{Clev1}) and (\ref{gammalev1}) to obtain
\be
C_{\alpha_1\alpha_2}^{\alpha\!,\,L_{-1}}
\gamma_{\vec\alpha\vec\alpha_3\vec\alpha_4}(L_{-1}) +
C_{\alpha_1\alpha_2}^{\alpha\!,\,W_{-1}}
\gamma_{\vec\alpha\vec\alpha_3\vec\alpha_4}(W_{-1}) =
-(\alpha\alpha_3+\beta\beta_3)
\frac{\Big(6\alpha\beta\alpha_1 + 3(\alpha^2-\beta^2)\beta_1\Big)}
{3\beta(3\alpha^2-\beta^2)} + \\ +
3\Big(-(\alpha^2-\beta^2)\alpha_3 + 2\alpha\beta\beta_3\Big)
\frac{\alpha_1\beta-\alpha\beta_1}{3\beta(3\alpha^2-\beta^2)}
= -\alpha_1\alpha_3-\beta_1\beta_3 = -\vec\alpha_1\vec\alpha_3
\ee
identically in $\alpha$ and $\beta$, and
in accordance with (\ref{frefco}).

\subsubsection{Switching on $c\neq 2$}

We do not fully repeat calculation in this case, because it adds
nothing new. Of importance are the deformations of operators:
\be
T = \frac{(\p\phi_1)^2+(\p\phi_2)^2}{2} + Q\p^2\phi_2,\\
W = \p\phi_1\Big((\p\phi_1)^2-3(\p\phi_2)^2\Big)
- \frac{3Q}{2}\Big(\p \phi_1\p^2\phi_2+3\p\phi_2\p^2\phi_1\Big)
- \frac{3Q^2}{2}\p^3\phi_2
\ee
The central charge is $c=2(1-6Q^2)$, and operator product expansion
(\ref{TWope}) remains intact.
Further,
\be
T(x)\, \no e^{\vec\alpha\vec\phi(0)}\no  \
= \frac{\alpha^2+\beta^2-2Q\beta}{2x^2}\no e^{\vec\alpha\vec\phi(0)}\no \ +
\frac{1}{x}\no \left(\p e^{\vec\alpha\vec\phi}\right)(0)\no \
+ \sum_{k\geq 0} x^k     \no \left(\frac{\p^k T}{k!}
+ \frac{\vec\alpha \p^{k+2}\vec\phi}{(k+1)!}\right)
e^{\vec\alpha\vec\phi}(z_2)\no
\ee
so that
\be
L_n\, \no e^{\vec\alpha\vec\phi}\no \ = 0, \ \ \ {\rm for}\ n>0,\\
L_0\, \no e^{\vec\alpha\vec\phi}\no \ = \Delta_{\alpha,\beta}
\no e^{\vec\alpha\vec\phi}\no \ =
\frac{\alpha^2+\beta^2-2Q\beta}{2}
\no e^{\vec\alpha\vec\phi}\no \ =
\frac{\alpha^2+\tilde\beta^2-Q^2}{2}
\no e^{\vec\alpha\vec\phi}\no  \\
L_{-1}\, \no e^{\vec\alpha\vec\phi}\no \ =\
\no \!(\alpha\p\phi_1 + \beta\p\phi_2)
e^{(\alpha\phi_1+\beta\phi_2)}\no  \\
L_{-2}\, \no e^{\vec\alpha\vec\phi}\no \ =
\no \!\left(\frac{1}{2} (\p\phi_1)^2 + \frac{1}{2}(\p\phi_2)^2 +
\alpha\p^2\phi_1 + (\beta+Q)\p^2\phi_2\right)
e^{(\alpha\phi_1+\beta\phi_2)}\no
\\
L_{-1}^2\, \no e^{\vec\alpha\vec\phi}\no \ =\
\no \!\Big(\alpha^2(\p\phi_1)^2 + 2\alpha\beta\p\phi_1\p\phi_2
+ \beta^2(\p\phi_2)^2
+ \alpha\p^2\phi_1 + \beta\p^2\phi_2\Big)
e^{(\alpha\phi_1+\beta\phi_2)}\no  \\
L_{-3}\, \no e^{\vec\alpha\vec\phi}\no \ =
\no \!\left(\p\phi_1\p^2\phi_1 + \p\phi_2\p^2\phi_2 +
\frac{\alpha}{2}\p^3\phi_1 + (\frac{\beta}{2}+Q)\p^3\phi_2\right)
e^{(\alpha\phi_1+\beta\phi_2)}\no
\label{L-expQ}
\ee
Similarly,
\be
W_n\, \no e^{\vec\alpha\vec\phi}\no \ = 0, \ \ \ {\rm for}\ n>0,\\
W_0\, \no e^{\vec\alpha\vec\phi}\no \ = w_{\alpha,\beta}
\no e^{\vec\alpha\vec\phi}\no \ =
\alpha(\alpha^2-3\beta^2+6Q\beta -3Q^2)
\no e^{\vec\alpha\vec\phi}\no \ =
\alpha(\alpha^2-3\tilde\beta^2)
\no e^{\vec\alpha\vec\phi}\no  \\
W_{-1}\, \no e^{\vec\alpha\vec\phi}\no \ =\
\no \!3\Big((\alpha^2-\beta^2+\frac{1}{2}Q\beta)\p\phi_1 -
\frac{1}{2}\alpha(4\beta-3Q) \p\phi_2\Big)
e^{(\alpha\phi_1+\beta\phi_2)}\no \\
W_{-2}\, \no e^{\vec\alpha\vec\phi}\no \ =
\ \no 3\Big(\alpha(\p\phi_1)^2 -2\beta \p\phi_1\p\phi_2 - \alpha(\p\phi_2)^2
+ (\alpha^2-\beta^2-Q\beta)\p^2\phi_1 - \alpha(2\beta-Q) \p^2\phi_2\Big)
e^{(\alpha\phi_1+\beta\phi_2)}\no
\label{W-expQ}
\ee
It is often convenient to use shifted variables
$\boxed{\tilde\beta = \beta-Q}$, where the both eigenvalues are simple:
\be
\Delta = \frac{\alpha^2+\tilde\beta^2-Q^2}{2},\ \ \ \
w = \alpha(\alpha^2-3\tilde\beta^2)
\ee
In this case the selection rule (conservation law)
in an $n$-point correlator of exponentials is
\be
\sum_{i=1}^n \alpha_i = 0,\ \ \ \sum_{i=1}^n \tilde\beta_i = (2-n)Q
\ee
Repeated application of $W$-operator gives
\be
W_1W_{-1} \, \no e^{\vec\alpha\vec\phi}\no \ = 9(\alpha^2+\beta^2-2Q\beta)
\left(\alpha^2+\beta^2-2Q\beta + \frac{3Q^2}{4}\right)
\no e^{\vec\alpha\vec\phi}\no \ = \frac{9D\Delta}{2} \no e^{\vec\alpha\vec\phi}\no \ =
\frac{9D}{2} L_{0} \no e^{\vec\alpha\vec\phi}\no
\label{W1W-1}
\ee
Here
\be D = 4\left(\alpha^2+\beta^2-2Q\beta + \frac{3Q^2}{4}\right) =
8\left(\Delta_{\alpha,\beta} + \frac{3Q^2}{8}\right) \label{Ddef}
\ee is a peculiar quantity in the theory of $W^{(3)}$
algebra.\footnote{ The factor $8$ in the definition of $D$ is
related to our choice of normalization for $\vec\alpha$. In
alternative normalization, accepted also in \cite{mmAGT}, with
$\alpha$ and $Q$ multiplied by $\sqrt{2}$ while $W$ and $w$
simultaneously divided by $\sqrt{8}$, one would get instead of
(\ref{Ddef}) \be\label{Ddef1} D= \varkappa
\left(\Delta-\frac{1}{5}\right) + \frac{1}{5}, \ \ \ \ \ \varkappa =
\left(1-\frac{15Q^2}{4}\right)^{-1}, \ \ \ c=2(1-12Q^2) \ee In this
case the $\varkappa$-dependent factor is also included into the
definition of the $W^{(3)}$ operator: $W\rightarrow \sqrt{\kappa}W$, so
that the eigenvalues of the two operators are $\Delta =
\alpha^2+\tilde\beta^2-Q^2$ and $w =
\sqrt{\varkappa}\alpha(\alpha^2-3\tilde\beta^2)$. With this choice
$D=\Delta$ when $Q=0$ and $c=2$. It is also natural from the point
of view of the AGT relation: in this normalization $Q$ coincides with
the parameter $\epsilon=\epsilon_1+\epsilon_2$  of Nekrasov
functions, $Q=\epsilon$. }
Next,
\be
W_0W_{-1} \! \no e^{\vec\alpha\vec\phi}\no \ =
3\alpha\Big(\alpha^2-3\beta^2+6Q\beta-3Q^2\Big)
\left((\alpha^2-\beta^2+\frac{1}{2}Q\beta)\p\phi_1 -
{2}\alpha\big(\beta-\frac{3Q}{4}\big)\p\phi_2\right) + \\
+ 18\Big(\alpha^2+\beta^2-2Q\beta+\frac{3Q^2}{4}\Big)
(\alpha\p\phi_1+\beta\p\phi_2)
=\left(wW_{-1} + \frac{9D}{2} L_{-1}\right) \no e^{\vec\alpha\vec\phi}\no
\label{W0W-1}
\ee
where $D$ is the same quantity (\ref{Ddef}) that appeared in (\ref{W1W-1}).
Actually, these relations (\ref{W1W-1}) and (\ref{W0W-1}) remain valid
in generic theory with the $W^{(3)}$ symmetry:
\be
W_0W_{-1} \longrightarrow w W_{-1} + \frac{9D}{2}L_{-1},
\ \ \ \ \ \ \ \ \ \ \ \ \ \
W_1W_{-1} \longrightarrow \frac{9D}{2}L_0,
\label{WWpri}
\ee
where arrow means that the relation is true, if operators act on a primary.
Finally,
\be
L_{-1}W_{-1}\, \no e^{\vec\alpha\vec\phi}\no \ =
3\Big( \alpha(\alpha^2-\beta^2+\frac{1}{2}Q\beta)(\p\phi_1)^2
- \big((\alpha^2+\beta^2)\beta -\frac{Q}{2}(3\alpha^2+\beta^2)\big)
\p\phi_1\p\phi_2
- \frac{1}{2}\alpha\beta(4\beta-3Q) (\p\phi_2)^2 +
\\
+(\alpha^2-\beta^2+\frac{1}{2}Q\beta)\p^2\phi_1 -
\frac{1}{2}\alpha(4\beta-3Q) \p^2\phi_2\Big)
e^{(\alpha\phi_1+\beta\phi_2)}\no  \
= \p \left(W_{-1}\no e^{\vec\alpha\vec\phi}\no \right)
\label{L-1W-1}
\ee
and
\be
\frac{1}{9}W_{-1}^2 \, \no e^{\vec\alpha\vec\phi}\no \ =
\left(\alpha^2-\beta^2+\frac{1}{2}Q\beta\right)
\left(\alpha^2-\beta^2+1+\frac{1}{2}Q\beta\right)
(\p\phi_1)^2 - \\
-\Big(4\alpha\beta(\alpha^2-\beta^2-1) + Q\alpha(-3\alpha^2+5\beta^2+3)
-\frac{3}{2}Q^2\alpha\beta\Big)\p\phi_1\p\phi_2+\\
+\Big(4\alpha^2\beta^2-\alpha^2+\beta^2 -Q\beta(6\alpha^2+\frac{1}{2})
+ \frac{9}{4}Q^2\alpha^2\Big)
(\p\phi_2)^2 + \\
+ \Big(2\alpha(\alpha^2+\beta^2) - Q\alpha\beta -\frac{3}{4}Q^2\alpha\Big)
\p^2\phi_1
+ \Big(2\beta(\alpha^2+\beta^2) - \frac{1}{2}Q(\alpha^2+7\beta^2)
+\frac{5}{4}Q^2\beta\Big)
\p^2\phi_2
\label{W-1sq}
\ee
Evaluation of correlators of these operators with exponentials is
straightforward: one simply substitutes field derivatives by
the corresponding powers of $\alpha$-parameters from the
counterpart exponential. Thus, one can easily check that the relations
from s.\ref{3PF} are, indeed, correct in the free field model.

\subsection{$\bar\Gamma$ vertices, structure constants and
Shapovalov matrix}

We now proceed to a discussion of $\bar\Gamma$ vertices.
This time the check is more tedious: directly defined in
the free field model are only the structure constants $C$.
They are related to the $\bar\Gamma$ vertices through the Shapovalov
matrix, which is, however, a much better studied and, hence,
reliable object. It depends only on the $\alpha$-parameters
of intermediate state. Moreover, in the free field model its
determinant (Kac determinant) factorizes nicely according to the
famous rule $\alpha = m\alpha_+ + n\alpha_-$, which lies
in the basis of the screening-operator approach to building
the CFT conformal blocks. As we mentioned in the Introduction,
this subject has a direct relation to the AGT conjecture \cite{AGT},
but we leave it beyond the scope of the present discussion.

Our general formulas for the $\bar\Gamma$-type vertices
are related to the structure constants evaluated for
the free field model in the previous subsection through
the Shapovalov matrix $Q$.
Therefore, in order to compare eqs.(\ref{barvirins}),
(\ref{viriter}) and (\ref{barWcor1})
with (\ref{Clev11f}), (\ref{Clev21f}) and
(\ref{Clev1}), one needs also
explicit expressions for $Q$. Fortunately, this is
a simple part of the problem: the Shapovalov matrix is rather easy to
calculate, if knowing the commutation relations for
the chiral algebra.

\subsubsection{One field, level one}

This is a very simple case. The matrix is $1\times 1$,
and its only element is
\be
\langle L_{-1}V_\alpha | L_{-1} V_\alpha \rangle\
\stackrel{(\ref{Hermi})}{=}\
\langle V_\alpha | L_1L_{-1} V_\alpha \rangle\
\stackrel{(\ref{virc})}{=} \
\langle V_\alpha | \Big(L_{-1}L_{1} + 2L_0\Big) V_\alpha \rangle\
\stackrel{(\ref{L-exp})}{=} \
2\Delta_\alpha \langle V_\alpha | V_\alpha \rangle\
= 2\Delta_\alpha H_{\alpha\alpha} = 2\Delta_\alpha
\label{Sha1f1l}
\ee
Here and in what follows we put
$H_{\alpha\alpha}\equiv \langle V_\alpha | V_\alpha \rangle\ = 1$,
because it appears as a common factor in all formulas.

Now one combines this with the basic relation (\ref{barGC})
and with the explicitly evaluated structure constant (\ref{Clev11f})
in the model of free fields, and obtains that, in this model,
\be
\bar\gamma_{\alpha_1\alpha_2;\alpha}(L_{-1})
= C_{\alpha_1\alpha_2}^{\alpha,L_{-1}}
\langle L_{-1}V_\alpha | L_{-1} V_\alpha \rangle\
\stackrel{(\ref{Clev11f})\&(\ref{Sha1f1l})}{=} \
\frac{\alpha_1}{\alpha}\,\delta_{\alpha,\alpha_1+\alpha_2}\cdot
\frac{2\alpha(\alpha-2Q)}{2} = \alpha_1(\alpha_1+\alpha_2-2Q)
\ee
in perfect agreement with (\ref{barvirins}), which
predicts for this case
\be
\bar\gamma_{\alpha_1\alpha_2;\alpha}(L_{-1})
\ \stackrel{(\ref{barvirins})}{=}\
\Delta_\alpha + \Delta_1-\Delta_2
\ \stackrel{\alpha=\alpha_1+\alpha_2}{=}\
\frac{(\alpha_1+\alpha_2)^2-2Q(\alpha_1+\alpha_2)
+ \alpha_1^2 -2Q\alpha_1 - (\alpha_2^2-2Q\alpha_2)}{2}=\\
= \alpha_1(\alpha_1+\alpha_2-2Q)
\label{a1a20}
\ee

\subsubsection{A check of relation (\ref{L-1V1})}

One can also check in the free-field model the other important relation,
(\ref{L-1V1}):
\be
\langle V_{\check\alpha}\ |\ (L_{-1}V_1)(1)\ V_2(0)\rangle \ =
\Big(\Delta_{\check\alpha} - \Delta_1-\Delta_2\Big)
\langle V_{\check\alpha}\ |\ V_1(1) V_2(0)\rangle
\label{L-1V1a}
\ee
Take $V_1 =\ \no e^{\alpha_1\phi}\no \ $ and $V_2=\ \no e^{\alpha_2\phi}\no \ $
Then, $L_{-1}V_1 = \ \no \alpha_1\p\phi e^{\alpha_1\phi}\no \ = \p V_1$
and the operator product expansion (\ref{ope}) in this case is simply
\be
\no \alpha_1\p\phi e^{\alpha_1\phi}(z)\no \ \no e^{\alpha_2\phi}(0)\no \
= z^{\alpha_1\alpha_2-1}\left(\alpha_1\alpha_2
\no e^{(\alpha_1+\alpha_2)\phi}(0)\no
+ z \no \alpha_1(1+\alpha_1\alpha_2)\p\phi
e^{(\alpha_1+\alpha_2)\phi}(0)\no  + \ldots\right)
\ee
Note that the first term at the r.h.s. was originally
$\alpha_1\alpha_2\no e^{(\alpha_1\phi(z)+\alpha_2\phi(0))}\no $
and its expansion in powers of $z$ makes all the terms in the
series non-vanishing and produces a correction to the order-$z$
term as well. Denoting also $V_\alpha = \ \no e^{\alpha\phi}\no \ $
with $\alpha=\alpha_1+\alpha_2$ so that
$L_{-1}V_\alpha =  \alpha \no \p\phi e^{\alpha\phi}\no \ $ one obtains
\be
(L_{-1}V_1)(z)\ V_2(0) = z^{\alpha_1\alpha_2-1}\left(
\alpha_1\alpha_2 V_\alpha(0) +
z \frac{\alpha_1(1+\alpha_1\alpha_2)}{\alpha_1+\alpha_2}
L_{-1}V_\alpha(0) + \ldots\right)
\ee
and
\be
\langle V_{\check\alpha}\ |\ (L_{-1}V_1)(1)\ V_2(0)\rangle \ =
\alpha_1\alpha_2 \langle V_{\check\alpha}\ |\ V_\alpha(0)\rangle\
+ \frac{\alpha_1(1+\alpha_1\alpha_2)}{\alpha_1+\alpha_2}
\langle V_{\check\alpha}\ |\ L_{-1}V_\alpha(0)\rangle\ + \ldots
\ee
Since the operators at the r.h.s. are taken at point $0$, the
matrix elements are those of the Shapovalov matrix, which we already
know up to level one.
Substituting $V_\alpha$ and $L_{-1}V_\alpha$ for $V_{\check\alpha}$,
one gets
\be
\langle V_{\alpha}\ |\ (L_{-1}V_1)(1)\ V_2(0)\rangle \ =
\alpha_1\alpha_2\delta_{\alpha,\alpha_1+\alpha_2}, \\
\langle L_{-1}V_{\alpha}\ |\ (L_{-1}V_1)(1)\ V_2(0)\rangle \ =
\frac{\alpha_1(1+\alpha_1\alpha_2)}{\alpha_1+\alpha_2}
\langle L_{-1}V_{\alpha}\ |\ L_{-1}V_{\alpha_1+\alpha_2}\rangle\
\stackrel{(\ref{Sha1f1l})}{=} \
\frac{\alpha_1(1+\alpha_1\alpha_2)}{\alpha_1+\alpha_2}\cdot
2\Delta_{\alpha} \delta_{\alpha,\alpha_1+\alpha_2}
\ee
the coefficients at the r.h.s. of these formulas being equal to
\be
\alpha_1\alpha_2 = \Delta_\alpha - \Delta_1-\Delta_2
= \frac{(\alpha_1+\alpha_2)^2
-2Q(\alpha_1+\alpha_2)+\alpha_1^2-2Q\alpha_1+\alpha_2^2-2Q\alpha_2}{2}
\label{a1a2}
\ee
and
\be
\frac{\alpha_1(1+\alpha_1\alpha_2)}{\alpha_1+\alpha_2}\cdot
\Big((\alpha_1+\alpha_2)^2-2Q(\alpha_1+\alpha_2)\Big)
= (1+\alpha_1\alpha_2)\alpha_1(\alpha_1+\alpha_2-2Q) = \\
\stackrel{(\ref{a1a2})\&(\ref{a1a20})}{=}(\Delta_\alpha + 1 - \Delta_1-\Delta_2)
(\Delta_\alpha+\Delta_1-\Delta_2)=
(\Delta_{\alpha,L_{-1}} - \Delta_1-\Delta_2)
(\Delta_\alpha+\Delta_1-\Delta_2)
\label{a1a21}
\ee
respectively, in full accordance with (\ref{L-1V1a}).
There is a product of two terms in the second line of (\ref{a1a21})
because (\ref{L-1V1a}) is a recursive relation:
\be
\langle L_{-1}V_{\alpha}\ |\ (L_{-1}V_1)(1)\ V_2(0)\rangle \
\stackrel{(\ref{L-1V1a})}{=}\
\Big(\Delta_{\alpha,L_{-1}} - \Delta_1-\Delta_2\Big)
\langle L_{-1}V_{\alpha}\ |\ V_1(1) V_2(0)\rangle\ =\\
\stackrel{(\ref{barvirins})}{=}\
\Big(\Delta_{\alpha,L_{-1}} - \Delta_1-\Delta_2\Big)
\Big(\Delta_{\alpha} + \Delta_1-\Delta_2\Big)
\langle V_{\alpha}\ |\ V_1(1) V_2(0)\rangle\
\ee
Note that all the signs in these formulas are absolutely essential
for the free-field calculation to get through -- this is how
we use free fields to validate (\ref{barvirins}) and (\ref{L-1V1}).

\subsubsection{One field, level two}

At level two the Shapovalov matrix is of the size $2\times 2$, with the elements
\be\label{ShL2}
\langle L_{-1}^{2}V_\alpha \Big| L_{-1}^{2} V_\alpha \rangle\
\stackrel{(\ref{Hermi})}{=}\
\langle V_\alpha \Big| L_1^2L_{-1}^2 V_\alpha \rangle\
\stackrel{(\ref{virc})}{=} \
\langle V_\alpha \Big| \Big(L_1L_{-1}L_{1}L_{-1} + 2L_1L_0L_{-1}\Big)
V_\alpha \rangle\ =\\
=\langle V_\alpha \Big| \Big(2L_1L_{-1}\Delta_\alpha + 2L_1L_{-1}+
2L_1L_{-1}\Delta_\alpha\Big) V_\alpha \rangle\
\stackrel{(\ref{L-exp})}{=} \
8\Delta_\alpha^2+4\Delta_\alpha \langle V_\alpha \Big| V_\alpha \rangle\
= 8\Delta_\alpha^2+4\Delta_\alpha
\\
\langle L_{-2}V_\alpha | L_{-2} V_\alpha \rangle\
\stackrel{(\ref{Hermi})}{=}\
\langle V_\alpha | L_2L_{-2} V_\alpha \rangle\
\stackrel{(\ref{virc})}{=} \
\langle V_\alpha \Big| \Big(L_{-2}L_{2} + \frac{1}{2} -
6Q^2 + 4L_0\Big) V_\alpha \rangle\
\stackrel{(\ref{L-exp})}{=} \
\frac{1}{2} - 6Q^2 + 4\Delta_\alpha
\\
\langle L_{-1}^{2}V_\alpha | L_{-2} V_\alpha \rangle\ =
\langle L_{-2}V_\alpha | L_{-1}^2 V_\alpha \rangle\
\stackrel{(\ref{Hermi})}{=}\
\langle V_\alpha | L_1L_1L_{-2} V_\alpha \rangle\
\stackrel{(\ref{virc})}{=} \
3\langle V_\alpha | L_1L_{-1} V_\alpha \rangle\ = 6\Delta_\alpha
\ee
Using this expression and the structure constants generalizing (\ref{Clev21f})
for $c\ne 1$, one can calculate the
$\bar\Gamma$-vertices:
\be
\bar\gamma_{\alpha_1\alpha_2;\alpha}(L_{-2})
= C_{\alpha_1\alpha_2}^{\alpha,L_{-2}}
\langle L_{-2}V_\alpha | L_{-2} V_\alpha \rangle\ +
\ C_{\alpha_1\alpha_2}^{\alpha,L_{-1}^2}\langle L_{-2}V_\alpha
| L_{-1}^2 V_\alpha \rangle\
=\\ =
\left(\frac{\alpha_{1}\alpha_{2}}{2\alpha(\alpha+Q)-1}\right)
\left(\frac{1}{2}+4\alpha(\alpha-2Q)-6Q^2\right)+
\left(\frac{\alpha_{1}(2\alpha_1(\alpha+Q)-1)}{2\alpha(2\alpha(\alpha+Q)-1)}\right)
\left(6\frac{\alpha(\alpha-2Q)}{2}\right)\ =\
\frac{\alpha_1}{2}(3\alpha_1+2\alpha_2-6Q)
\ee
and
\be
\bar\gamma_{\alpha_1\alpha_2;\alpha}(L_{-1}^2)
= C_{\alpha_1\alpha_2}^{\alpha,L_{-2}}
\langle L_{-1}^2V_\alpha | L_{-2} V_\alpha \rangle\ +
\ C_{\alpha_1\alpha_2}^{\alpha,L_{-1}^2}\langle L_{-1}^2V_\alpha
| L_{-1}^2 V_\alpha \rangle\
=\\ =
\left(\frac{\alpha_{1}\alpha_{2}}{2\alpha(\alpha+Q)-1}\right)
\left(6\frac{\alpha(\alpha-2Q)}{2}\right)+
\left(\frac{\alpha_{1}(2\alpha_1(\alpha+Q)-1)}{2\alpha(2\alpha(\alpha+Q)-1)}\right)
\left(4\frac{\alpha(\alpha-2Q)}{2}(2\frac{\alpha(\alpha-2Q)}{2}+1)\right)\ = \\ =
\ \alpha_1(\alpha_1^2+\alpha_1\alpha_2-2\alpha_1Q+1)(\alpha_1+\alpha_2-2Q)
\ee
which agree with (\ref{barvirins}) predicting in this case
\be
\bar\gamma_{\alpha_1\alpha_2;\alpha}(L_{-2})
\ \stackrel{(\ref{barvirins})}{=}\
\Delta_\alpha + 2\Delta_1-\Delta_2
\ \stackrel{\alpha=\alpha_1+\alpha_2}{=}\
\frac{(\alpha_1+\alpha_2)^2-2Q(\alpha_1+\alpha_2)
+ 2\alpha_1^2 -4Q\alpha_1 - (\alpha_2^2-2Q\alpha_2)}{2}=\\
= \frac{\alpha_1}{2}(3\alpha_1+2\alpha_2-6Q)
\ee
and
\be
\bar\gamma_{\alpha_1\alpha_2;\alpha}(L_{-1}^2)
\ \stackrel{(\ref{barvirins})}{=}\
(\Delta_\alpha + \Delta_1-\Delta_2)(\Delta_\alpha + \Delta_1-\Delta_2+1)
\ \stackrel{\alpha=\alpha_1+\alpha_2}{=} \\=
\frac{(\alpha_1+\alpha_2)^2-2Q(\alpha_1+\alpha_2)
+ \alpha_1^2 -2Q\alpha_1 - (\alpha_2^2-2Q\alpha_2)}{2}\times\\\times
\left(\frac{(\alpha_1+\alpha_2)^2-2Q(\alpha_1+\alpha_2)
+ \alpha_1^2 -2Q\alpha_1 - (\alpha_2^2-2Q\alpha_2)}{2}+1\right)=\\
= \alpha_1(\alpha_1^2+\alpha_1\alpha_2-2\alpha_1Q+1)(\alpha_1+\alpha_2-2Q)
\ee

\subsubsection{Two fields, level one}

This time the Shapovalov matrix is also $2\times 2$, but of course
very different from (\ref{ShL2}). Moreover, this time
to calculate it, one needs commutation relations for the $W$ algebra. The result reads
\be
Q=\left(\begin{array}{cc}
2\Delta & 3w \\
3w & 9D\Delta/2
\end{array}\right)
\label{SF1}
\ee
Again, one can calculate the $\bar\Gamma$-vertices using this matrix and
generalizing (\ref{Clev1}) for $c\ne 2$:
\be
\bar\gamma_{\alpha_1\alpha_2;\alpha}(L_{-1})
= C_{\alpha_1\alpha_2}^{\alpha,L_{-1}}
\langle L_{-1}V_\alpha | L_{-1} V_\alpha \rangle\ +
\ C_{\alpha_1\alpha_2}^{\alpha,W_{-1}}\langle L_{-1}V_\alpha
| W_{-1} V_\alpha \rangle\
=\\ =
\left(\frac{-4\alpha_1\alpha\beta+3\alpha_1\alpha Q-2\beta_1\alpha^2+2\beta_1\beta^2-
\beta_1Q\beta}{-6\alpha^2\beta+3\alpha^2Q+2\beta^3-Q\beta^2}\right)
\left(2\frac{\alpha(\alpha-2Q)}{2}\right)+\\+
\left(\frac{2(\alpha\beta_1-\alpha_1\beta)}{3(-6\alpha^2\beta+3\alpha^2Q+2\beta^3-
Q\beta^2)}\right)
\left(3\alpha(\alpha^2-3\beta^2+6Q\beta-3Q^2)\right)\ =\\=
\ \frac{\alpha^2+\beta^2+\alpha_1^2+\beta_1^2-\alpha_2^2-\beta_2^2-2Q\beta-2Q\beta_1+
2Q\beta_2}{2}
\ee
which agree with (\ref{barvirins})
\be
\bar\gamma_{\alpha_1\alpha_2;\alpha}(L_{-1})
\ \stackrel{(\ref{barvirins})}{=}\
\Delta_\alpha + \Delta_1-\Delta_2
\ \stackrel{\alpha=\alpha_1+\alpha_2}{=}\ \frac{\alpha^2+\beta^2+\alpha_1^2+
\beta_1^2-\alpha_2^2-\beta_2^2-2Q\beta-2Q\beta_1+2Q\beta_2}{2}
\ee
At the same time, the second vertex allows one to calculate $\langle V_{\check\alpha}
|(W_{-1}V_1)(1)\ V_2(0) \rangle$. Indeed, from the Shapovalov matrix and the
structure constants one obtains
\be
\bar\gamma_{\alpha_1\alpha_2;\alpha}(W_{-1})
= C_{\alpha_1\alpha_2}^{\alpha,L_{-1}}
\langle W_{-1}V_\alpha | L_{-1} V_\alpha \rangle\ +
\ C_{\alpha_1\alpha_2}^{\alpha,W_{-1}}\langle W_{-1}V_\alpha | W_{-1}
V_\alpha \rangle\
=\\ =
\left(\frac{-4\alpha_1\alpha\beta+3\alpha_1\alpha Q-2\beta_1\alpha^2+2\beta_1\beta^2-
\beta_1Q\beta}{-6\alpha^2\beta+3\alpha^2Q+2\beta^3-Q\beta^2}\right)
\left(3\alpha(\alpha^2-3\beta^2+6Q\beta-3Q^2)\right)+\\+
\left(\frac{2(\alpha\beta_1-\alpha_1\beta)}{3(-6\alpha^2\beta+3\alpha^2Q+2\beta^3-
Q\beta^2)}\right)
\left(\frac{9\alpha(\alpha-2Q)}{4}\left(8\frac{\alpha(\alpha-2Q)}{2}+3Q^2
\right)\right)\ = \\ =
\ \frac{6\alpha_1\alpha^2-12\beta_1\beta\alpha-6\beta^2\alpha_1-18\alpha_1Q^2+
15Q\alpha\beta_1+21Q\alpha_1\beta}{2}=\\=w+2w_1-w_2+{3\over 2}\alpha\beta_1 Q-
6\alpha_1\beta_1 Q-6\beta\beta_1\alpha_1+9\alpha_1\beta_1^2+{9\over 2}
\alpha_1\beta Q+3\alpha\alpha_1^2-3\alpha_1^3-3\alpha\beta_1^2=\\
=w-w_1-w_2+(3(\alpha_1^2-\beta_1^2)\alpha-6\alpha_1\beta_1\beta)+
{3\over 2}Q(\alpha\beta_1+8\alpha_1\beta_1+3\alpha_1\beta-6\alpha_1 Q)
\ee
Now, using (\ref{barWcor1}) and (\ref{51}), one gets
\be
\langle V_{\check\alpha} |(W_{-1}V_1)(1)\ V_2(0) \rangle\ =
3\left[{\alpha\beta_1 Q\over 2}-
2\alpha_1\beta_1 Q-2\beta\beta_1\alpha_1+3\alpha_1\beta_1^2+{3\over 2}
\alpha_1\beta Q+\alpha\alpha_1^2-\alpha_1^3-\alpha\beta_1^2\right]
\langle V_{\check\alpha} |V_1(1)\ V_2(0) \rangle\
\\
\langle V_{\check\alpha} |V_1(1)\ (W_{-1}V_2)(0) \rangle\
=3\left[((\alpha_1^2-\beta_1^2)\alpha-2\alpha_1\beta_1\beta)+
{Q\over 2}(\alpha\beta_1+8\alpha_1\beta_1+3\alpha_1\beta-6\alpha_1 Q)\right]
\langle V_{\check\alpha} |V_1(1)\ V_2(0) \rangle\
\ee
The latter expression coincides with the $\Gamma$-vertex
$\langle V_{\check\alpha} (0)(W_{-1}V_1)(1)\ V_2(\infty) \rangle$ in
(\ref{Wlev13}) at $Q=0$. For the
special state $V_1(1)$, (\ref{ss}) one obtains
\be
\langle V_{\check\alpha} |(W_{-1}V_1)(1)\ V_2(0) \rangle\ ={3w_1\over 2\Delta_1}
\Big(\Delta-\Delta_1-\Delta_2\Big), \ \ \ \ \ \ \ \ \
\langle V_{\check\alpha} |V_1(1)\ (W_{-1}V_2)(0) \rangle\ = -{3w_1\over 2\Delta_1}
\Big(\Delta+\Delta_1-\Delta_2\Big)
\ee
The first expression is in complete agreement with (\ref{specW-1}), while
the second one again coincides with $\langle V_{\check\alpha} (0)(W_{-1}V_1)(1)
\ V_2(\infty) \rangle$, see (\ref{specw-1}).

\subsubsection{Two fields, level two}

In this case the Shapovalov matrix is much more involved:

$$
\hspace{-1cm}\begin{array}{|c||c|c|c|c|c|}
\hline
&&&&&\\
Q_{\alpha}({\cal Y},{\cal Y}')
& L_{-2} V_{\alpha} &  L_{-1}^2 V_{\alpha} & L_{-1}W_{-1} V_{\alpha}
& W_{-2} V_{\alpha} & W_{-1}^2 V_{\alpha} \\
&&&&&\\
\hline\hline
&&&&&\\
L_{-2} V_{\alpha}&4\Delta+1-6Q^2&6\Delta&9w&6w&\frac{45D\Delta}{2}\\
&&&&&\\
\hline
&&&&&\\
 L_{-1}^2 V_{\alpha}&6\Delta&4\Delta(2\Delta+1)&6w(2\Delta+1)&12w
 &9(3D\Delta+2w^2)\\
&&&&&\\
\hline
&&&&&\\
L_{-1}W_{-1} V_{\alpha}&9w&6w(2\Delta+1)&9(D\Delta^2 + D\Delta + w^2)
&18D\Delta&\frac{27Dw}{2}(2\Delta+3)\\
&&&&&\\
\hline
&&&&&\\
W_{-2} V_{\alpha}&6w&12w&18D\Delta
&72\Delta\left(\Delta+1-\frac{3Q^2}{2}\right)
&108w\left(3\Delta+1-\frac{3Q^2}{4}\right)\\
&&&&&\\
\hline
&&&&&\\
W_{-1}^2 V_{\alpha}&\frac{45D\Delta}{2}&9(3D\Delta+2w^2)
&\frac{27Dw}{2}(2\Delta+3)& 108w\left(3\Delta+1-\frac{3Q^2}{4}\right)
&\frac{81}{4}
D^2\Delta(2\Delta+1)\ +\\
&&&&&162 \Big(D\Delta(\Delta+1)+4w^2
\Big)\\
&&&&&\\
\hline
\end{array}
\label{SF2}
$$

\bigskip

\noindent
However, using formulas of s.9.1.5 for the structure constants, one can
repeat the calculations of the previous subsections in this case too. We do not
list the formulas here because of their complexity.

Note that in $\alpha$-parametrization the determinant of the Shapovalov matrix
(Kac determinant) factorizes, as usual: 
\be
\det Q = \frac{3^8}{4}\left\{(4\tilde\beta^2-Q^2)
\Big((\tilde\beta+Q)^2-3\alpha^2\Big)\Big((\tilde\beta-Q)^2-3\alpha^2\Big)
\right\}^2
\Big(2\tilde\beta^2+3Q\tilde\beta+Q^2-1\Big)
\Big(2\tilde\beta^2-3Q\tilde\beta+Q^2-1\Big)
\cdot\nn\\ \cdot
\Big(4+12\tilde\beta Q^3+6\tilde\beta^3 Q-6\tilde\beta Q-18\alpha^2\tilde\beta Q
-15\alpha^2 Q^2-12\alpha^2-4\tilde\beta^2-4Q^2-6\alpha^2\tilde\beta^2
+13\tilde\beta^2Q^2+\tilde\beta^4+4Q^4+9\alpha^4\Big)\cdot\nn\\
\Big(4-12\tilde\beta Q^3-6\tilde\beta^3 Q+6\tilde\beta Q+18\alpha^2\tilde\beta Q
-15\alpha^2 Q^2-12\alpha^2-4\tilde\beta^2-4Q^2-6\alpha^2\tilde\beta^2
+13\tilde\beta^2 Q^2+\tilde\beta^4+4Q^4+9\alpha^4\Big)
\ee
See \cite{mmAGT} for further details.

\section{Summary for the needs of AGT considerations}

\subsection{Collection of recursion relations}

We list here a few concrete expressions from s.\ref{3PF},
which are used in \cite{MMMagt} and \cite{mmAGT}
for the study of AGT relations \cite{AGT} between
conformal blocks and Nekrasov functions.
These are recurrent relations, directly applicable to primaries,
and should be iterated if applied to descendants.
They are different the $\bar\Gamma$- and $\Gamma$-type triple vertices.
The fields $V_1,V_2,V_3,V_4$ are assumed below to be Virasoro and $W^{(3)}$
primaries respectively.

\bigskip

{\underline{\bf Virasoro sector:}}

\be
\boxed{
\langle L_{-n} V_{\check\alpha}\ |\ V_1(1)V_2(0) \rangle\
\stackrel{(\ref{barvirins})}{=}\
\Big(\Delta_{\check\alpha}  + n\Delta_1  - \Delta_2\Big)
\langle V_{\check\alpha}\ |\ V_1(1)V_2(0) \rangle,} \ \ \ \ n>0
\label{barvirinsC}
\ee
\be
\boxed{
\langle (L_{-n} V_{\check\alpha})(0)\ V_3(1)V_4(\infty) \rangle\
\stackrel{(\ref{virins})}{=}\
\Big(\Delta_{\check\alpha}  + n\Delta_3  - \Delta_4\Big)
\langle V_{\check\alpha}(0)\  V_3(1)V_4(\infty) \rangle,} \ \ \ \ n>0
\label{virinsC}
\ee
Also useful are
\be
\langle V_{\check\alpha}\ |\ (L_{-1}V_1)(1)\ V_2(0)\rangle \
\stackrel{(\ref{L-1V1})}{=}\
\Big(\Delta_{\check\alpha} - \Delta_1-\Delta_2\Big)
\langle V_{\check\alpha}\ |\ V_1(1) V_2(0)\rangle
\label{L-1V1C}
\ee
and
\be
\langle V_{\check\alpha}(0)\ (L_{-1}V_3)(1)\ V_4(\infty)\rangle\
\stackrel{(\ref{L-1V3})}{=}\
-\Big(\Delta_{\check\alpha} +\Delta_3-\Delta_4\Big)
\langle V_{\check\alpha}(0)\ V_3(1)\ V_4(\infty)\rangle
\label{L-1VC}
\ee
The last two relations hold when $V_1$ and $V_3$ are arbitrary operators,
not obligatory primaries.

\bigskip

{\underline{\bf $W^{(3)}$-operator sector:}}

\fr{
\langle W_{-n}V_{\check\alpha} | V_1(1)\ V_2(0) \rangle\
\stackrel{(\ref{barWcor1})}{=}\
\langle W_{0}V_{\check\alpha} | V_1(1)\ V_2(0) \rangle +
\left(\frac{n(n+3)}{2}w_1-w_2\right)
\langle V_{\check\alpha} | V_1(1)\ V_2(0) \rangle
+\\+
n\langle V_{\check\alpha}  | (W_{-1}V_1)(1)\ V_2(0)\rangle,
\ \ \ \ \ \ n>0
\label{barWcor1C}}
\fr{
\langle (W_{-n}V_{\check\alpha})(0)\ V_3(1)\ V_4(\infty) \rangle \
\stackrel{(\ref{Wcor3})}{=}\
\langle (W_{0}V_{\check\alpha})(0)\ V_3(1)\ V_4(\infty) \rangle \
+ \left( -\frac{n(n-3)}{2}w_3+w_4\right)
\langle V_{\check\alpha}(0)\ V_3(1)\ V_4(\infty) \rangle \ +\\
+ n\langle V_{\check\alpha}(0)\ (W_{-1}V_3)(1)\ V_4(\infty)\rangle,
\ \ \ \ \ \ n>0
\label{Wcor3C}}
Also of use are the two particular examples where no operators are
assumed to be primaries:
\be
\langle W_{-1}V_{\check\alpha} | V_1(1)\ V_2(0) \rangle \
\stackrel{(\ref{barWcor1np})}{=}\
\langle W_{0}V_{\check\alpha} | V_1(1)\ V_2(0) \rangle \
+ 2\langle V_{\check\alpha} | (W_{0}V_1)(1)\ V_2(0) \rangle \
-\ \langle V_{\check\alpha} | V_1(1)\ (W_{0}V_2)(0) \rangle \ +\\
+ \ \langle V_{\check\alpha} | (W_{-1}V_1)(1)\ V_2(0) \rangle \
+ \ \langle V_{\check\alpha} | (W_{1}V_1)(1)\ V_2(0) \rangle \
+ \ \langle V_{\check\alpha} | V_1(1)\ (W_{1}V_2)(0) \rangle
\label{barWcor1npC}
\ee
and
\be
\langle (W_{-1}V_{\check\alpha})(0)\ V_3(1)\ V_4(0) \rangle \
\stackrel{(\ref{Wcor3np})}{=}\
\langle W_{0}V_{\check\alpha}(0)\ V_3(1)\ V_4(0) \rangle \
+ \langle V_{\check\alpha}(0)\ (W_{0}V_3)(1)\ V_4(0) \rangle +\\
+ \langle V_{\check\alpha}(0)\ V_3(1)\ (W_{0}V_4)(0) \rangle \
+  \langle V_{\check\alpha}(0)\ (W_{-1}V_3)(1)\ V_4(0) \rangle \
- \ \langle V_{\check\alpha}(0)\ V_3(1)\ (W_{1}V_4)(0) \rangle
\label{Wcor3npC}
\ee
and two relations, analogous to (\ref{L-1V1C}) and (\ref{L-1VC}),
\be
\langle V_{\check\alpha}|(W_{-2}V_1)(1)\ V_2(0) \rangle\ =
\Big(\hat w_{\check\alpha}-w_1-w_2\Big)
\langle V_{\check\alpha}|V_1(1)\ V_2(0) \rangle\ -
2\langle V_{\check\alpha}|(W_{-1}V_1)(1)\ V_2(0) \rangle
\label{AABbarC}
\ee
and
\be
\langle V_{\check\alpha}(0)\ (W_{-2}V_3)(1)V_4(\infty) \rangle\ =
-\Big(\hat w_{\check\alpha}+w_3+w_4\Big)
\langle V_{\check\alpha}(0)\ V_3(1)V_4(\infty) \rangle\
- 2\langle V_{\check\alpha}(0)\ (W_{-1}V_3)(1)V_4(\infty) \rangle
\label{AABC}
\ee
Actually, for evaluation of conformal blocks in
\cite{mmAGT} one needs particular versions of these formulas.
On one hand, one needs them only at low levels, $n\leq 2$,
on the other hand, one needs them adjusted to imposing {\it the speciality
condition} on some of the states (which we choose to be $V_1$ and $V_3$).
This condition needs to be imposed on the vertex operators,
if one wants to unambiguously evaluate all conformal blocks
without specifying a conformal model. Otherwise, in the theory with the
$W^{(3)}$ chiral algebra our recursion relations allow one only to reduce
arbitrary triple vertices to the two sets
$\langle V_\alpha |\ (W_{-1}^k V_1)(1) \ V_2(0) \rangle$
and $\langle V_{\alpha}(0)\ (W_{-1}^k V_3)(1)\ V_4(\infty)\rangle$
which are arbitrary parameters, depending on the particular model.
{\it The speciality condition} expresses $W_{-1}V_{1,3}$
through $L_{-1}V_{1,3}$ and gets rid of the uncertainty for the
restricted set of conformal blocks: the $m$-point conformal blocks
with $m-2$ {\it special} external legs are unambiguously dictated
by the $W^{(3)}$ symmetry. Of course, in the given conformal model {\it all}
the conformal blocks are unambiguously defined, but they are not all
dictated by symmetry and, therefore, are pretty hard to evaluate --
of course, if the model is not that of the free fields.
In the remaining part of this section we convert the recursion relations
(\ref{barvirinsC})-(\ref{AABC}) into the form relevant for \cite{mmAGT}.

\subsection{$\bar\Gamma$-type vertices (bilinear forms)}

\subsubsection{Six simple cases}

Directly from (\ref{barvirinsC}) and (\ref{barWcor1C})
for three primaries $V_\alpha$, $V_1$, $W_2$ one obtains
\be
\boxed{
\langle L_{-1} V_{\alpha}\ |\ V_1(1)V_2(0) \rangle\
\stackrel{(\ref{barvirinsC})}{=}\
\Big(\Delta_{\alpha}  + \Delta_1  - \Delta_2\Big)
\langle V_{\alpha}\ |\ V_1(1)V_2(0) \rangle,}
\label{L-1C}
\ee
\be
\boxed{
\langle W_{-1}V_{\alpha} | V_1(1)\ V_2(0) \rangle\
\stackrel{(\ref{barWcor1C})}{=}\
\left(w_\alpha+2w_1-w_2\right)
\langle V_{\alpha} | V_1(1)\ V_2(0) \rangle
+ \underline{\langle V_{\check\alpha}  | (W_{-1}V_1)(1)\ V_2(0)\rangle}
}
\label{W-1C}
\ee
\be
\boxed{
\langle L_{-2} V_{\alpha}\ |\ V_1(1)V_2(0) \rangle\
\stackrel{(\ref{barvirinsC})}{=}\
\Big(\Delta_{\alpha}  + 2\Delta_1  - \Delta_2\Big)
\langle V_{\alpha}\ |\ V_1(1)V_2(0) \rangle,}
\label{L-2C}
\ee
\be
\boxed{
\langle W_{-2}V_{\check\alpha} | V_1(1)\ V_2(0) \rangle\
\stackrel{(\ref{barWcor1C})}{=}\
\left(w_\alpha +5w_1-w_2\right)
\langle V_{\check\alpha} | V_1(1)\ V_2(0) \rangle
+ 2\underline{\langle V_{\check\alpha}  | (W_{-1}V_1)(1)\ V_2(0)\rangle}
}
\label{W-2C}
\ee
In two iterations of (\ref{barvirinsC})
\be
\boxed{
\langle L_{-1}^2 V_{\alpha}\ |\ V_1(1)V_2(0) \rangle\
}
\stackrel{(\ref{barvirinsC})}{=}\
\Big((\Delta_{\alpha}+1)  + \Delta_1  - \Delta_2\Big)
\langle L_{-1}V_{\alpha}\ |\ V_1(1)V_2(0) \rangle\ = \\
\boxed{
\stackrel{(\ref{barvirinsC})}{=}\
\Big(\Delta_{\alpha}  + \Delta_1  - \Delta_2+1\Big)
\Big(\Delta_{\alpha}  + \Delta_1  - \Delta_2\Big)
\langle V_{\alpha}\ |\ V_1(1)V_2(0) \rangle,}
\label{L-1L-1C}
\ee
In two iterations from (\ref{barvirinsC}) and (\ref{W-1C})
\be
\boxed{
\langle L_{-1}W_{-1} V_{\alpha}\ |\ V_1(1)V_2(0) \rangle}\
\stackrel{(\ref{barvirinsC})}{=}\
\Big((\Delta_{\alpha}+1)  + \Delta_1  - \Delta_2\Big)
\langle W_{-1}V_{\alpha}\ |\ V_1(1)V_2(0) \rangle\ =\\
\boxed{
\stackrel{(\ref{W-1C})}{=}\
\Big(\Delta_{\alpha}  + \Delta_1  - \Delta_2+1\Big)
\Big(\left(w_\alpha+2w_1-w_2\right)
\langle V_{\alpha} | V_1(1)\ V_2(0) \rangle
+ \underline{\langle V_{\check\alpha}  | (W_{-1}V_1)(1)\ V_2(0)\rangle\Big)}
}
\label{L-1W-1C}
\ee
In the last equality we used the fact that dimension
$\Delta(W_{-1}V_\alpha) = \Delta_\alpha + 1$.
Underlined are the $\bar\Gamma$-vertices in the answers, which are different from
$\langle V_{\alpha} | V_1(1)\ V_2(0) \rangle$.

\subsubsection{The difficult case: $W_{-1}^2$}

For three primaries $V_\alpha$, $V_1$, $W_2$
and for $V_{\check\alpha} = W_{-1}V_\alpha$
it follows immediately from (\ref{barWcor1npC}):
\be
\boxed{\langle W_{-1}^2V_{\alpha} | V_1(1)\ V_2(0) \rangle = }\\
\stackrel{(\ref{barWcor1npC})}{=}\
\langle W_{0}W_{-1}V_{\alpha} | V_1(1)\ V_2(0) \rangle \
+ 2\langle W_{-1}V_{\alpha} | (W_{0}V_1)(1)\ V_2(0) \rangle \
-\ \langle W_{-1}V_{\alpha} | V_1(1)\ (W_{0}V_2)(0) \rangle \ + \\
+ \ \langle W_{-1}V_{\alpha} | (W_{-1}V_1)(1)\ V_2(0) \rangle \
+ \ \langle W_{-1}V_{\alpha} | (W_{1}V_1)(1)\ V_2(0) \rangle \
+ \ \langle W_{-1}V_{\alpha} | V_1(1)\ (W_{1}V_2)(0) \rangle =  \\
= \langle W_{0}W_{-1}V_{\alpha} | V_1(1)\ V_2(0) \rangle \
+ (2w_1-w_2)\langle W_{-1}V_{\alpha} | V_1(1)\ V_2(0) \rangle \
+ \ \langle W_{-1}V_{\alpha} | (W_{-1}V_1)(1)\ V_2(0) \rangle
\label{barWcor1np11}
\ee
The first term at the r.h.s. can be handled with the help of
(\ref{WWpri}), while (\ref{W-1C}) is directly applicable to the second term.
To handle the last term,
one applies the same (\ref{barWcor1npC})
but with non-primary $V_1\rightarrow W_{-1}V_1$ and
primary $V_{\check\alpha} = V_\alpha$:
\be
\langle W_{-1}V_{\alpha} | (W_{-1}V_1)(1)\ V_2(0) \rangle \ = \\
\stackrel{(\ref{barWcor1npC})}{=}\
\langle W_{0}V_{\alpha} | W_{-1}V_1(1)\ V_2(0) \rangle \
+ 2\langle V_{\alpha} | (W_{0}W_{-1}V_1)(1)\ V_2(0) \rangle \
-\ \langle V_{\alpha} | W_{-1}V_1(1)\ (W_{0}V_2)(0) \rangle \ +\\
+ \ \langle V_{\alpha} | (W_{-1}^2V_1)(1)\ V_2(0) \rangle \
+ \ \langle V_{\alpha} | (W_{1}W_{-1}V_1)(1)\ V_2(0) \rangle \
+ \ \langle V_{\alpha} | W_{-1}V_1(1)\ (W_{1}V_2)(0) \rangle\ =\\
= (w_\alpha-w_2)\langle V_{\alpha} | W_{-1}V_1(1)\ V_2(0) \rangle \
+ 2\langle V_{\alpha} | (W_{0}W_{-1}V_1)(1)\ V_2(0) \rangle \
+ \ \langle V_{\alpha} | (W_{-1}^2V_1)(1)\ V_2(0) \rangle \ + \\
+ \ \langle V_{\alpha} | (W_{1}W_{-1}V_1)(1)\ V_2(0) \rangle \
\label{barWcor1np1-1}
\ee
The next step is to substitute
\be
W_0W_{-1} \longrightarrow w W_{-1} + \frac{9D}{2}L_{-1},
\ \ \ \ \ \ \ \ \ \ \
W_1W_{-1} \longrightarrow \frac{9D}{2}L_0
\label{WWpri1}
\ee
from (\ref{WWpri}).
Arrows mean that these relations are true, if operators act on primaries,
and $D$ was defined in (\ref{Ddef}).

Then one gets from (\ref{barWcor1np11})
with substituted (\ref{barWcor1np1-1}):
\be
\langle W_{-1}^2V_{\alpha} | V_1(1)\ V_2(0) \rangle \ =
\frac{9D_\alpha}{2}\langle L_{-1}V_{\alpha} | V_1(1)\ V_2(0) \rangle \
+ (w_\alpha+ 2w_1-w_2)\langle W_{-1}V_{\alpha} | V_1(1)\ V_2(0) \rangle \
+ \\
+(w_\alpha+2w_1-w_2)\langle V_{\alpha} | W_{-1}V_1(1)\ V_2(0) \rangle \
+ 9D_1\langle V_{\alpha} | (L_{-1}V_1)(1)\ V_2(0) \rangle \
+ \ \langle V_{\alpha} | (W_{-1}^2V_1)(1)\ V_2(0) \rangle \ + \\
+ \frac{9D_1\Delta_1}{2} \langle V_{\alpha} | V_1(1)\ V_2(0) \rangle
\ee
It remains to substitute (\ref{W-1C}), (\ref{L-1C}) and (\ref{L-1V1C})
in order to get
\be
\langle W_{-1}^2V_{\alpha} | V_1(1)\ V_2(0) \rangle\ = \
(w_\alpha+ 2w_1-w_2)\Big(
(w_\alpha+2w_1-w_2)
\langle V_{\alpha} | V_1(1)\ V_2(0) \rangle
+ 2\underline{\langle V_{\alpha} | (W_{-1}V_1)(1)\ V_2(0)\rangle}
\Big) +
\\
+ \frac{9D_\alpha}{2}
\Big(\Delta_{\alpha}  + \Delta_1  - \Delta_2\Big)
\langle V_{\alpha} | V_1(1)V_2(0) \rangle
+ 9D_1
\Big(\Delta_{\alpha} - \Delta_1-\Delta_2\Big)
\langle V_{\alpha} | V_1(1) V_2(0)\rangle
+
\\
+ \frac{9D_1\Delta_1}{2} \langle V_{\alpha} | V_1(1)\ V_2(0)\rangle
+ \underline{\underline{
\langle V_{\alpha} | (W_{-1}^2V_1)(1)\ V_2(0) \rangle}}
\label{W-1W-1C}
\ee

\subsection{$\Gamma$-type vertices (linear forms or correlators)}

\subsubsection{Six simple cases}

Directly from (\ref{virinsC}) and (\ref{Wcor3C})
for three primaries $V_\alpha$, $V_1$, $W_2$ one obtains
\be
\boxed{
\langle (L_{-1} V_{\alpha})(0)\ V_3(1)V_4(\infty) \rangle\
\stackrel{(\ref{virinsC})}{=}\
\Big(\Delta_{\alpha}  + \Delta_3  - \Delta_4\Big)
\langle V_{\alpha}(0)\  V_3(1)V_4(\infty) \rangle,}
\label{l-1C}
\ee
\be
\boxed{
\langle (W_{-1}V_{\alpha})(0)\ V_3(1)\ V_4(\infty) \rangle \
\stackrel{(\ref{Wcor3C})}{=}\
\left(w_\alpha + w_3+w_4\right)
\langle V_{\alpha}(0)\ V_3(1)\ V_4(\infty) \rangle \
+ \underline{\langle V_{\alpha}(0)\ (W_{-1}V_3)(1)\ V_4(\infty)\rangle},
}
\label{w-1C}
\ee
\be
\boxed{
\langle (L_{-2} V_{\alpha})(0)\ V_3(1)V_4(\infty) \rangle\
\stackrel{(\ref{virinsC})}{=}\
\Big(\Delta_{\alpha}  + 2\Delta_3  - \Delta_4\Big)
\langle V_{\alpha}(0)\  V_3(1)V_4(\infty) \rangle,}
\label{l-2C}
\ee
\be
\boxed{
\langle (W_{-2}V_{\alpha})(0)\ V_3(1)\ V_4(\infty) \rangle \
\stackrel{(\ref{Wcor3C})}{=}\
\left(w_\alpha + w_3+w_4\right)
\langle V_{\alpha}(0)\ V_3(1)\ V_4(\infty) \rangle \
+ 2\underline{\langle V_{\alpha}(0)\ (W_{-1}V_3)(1)\ V_4(\infty)\rangle},
}
\label{w-2C}
\ee
In two iterations of (\ref{virinsC}):
\be
\boxed{
\langle (L_{-1}^2 V_{\check\alpha})(0)\ V_3(1)V_4(\infty) \rangle}\
\stackrel{(\ref{virinsC})}{=}\
\Big((\Delta_{\alpha}+1)  + \Delta_3  - \Delta_4\Big)
\langle L_{-1}V_{\alpha}(0)\  V_3(1)V_4(\infty) \rangle\ = \nn\\
\boxed{\stackrel{(\ref{virinsC})}{=}\
\Big(\Delta_{\alpha}  + \Delta_3  - \Delta_4+1\Big)
\Big(\Delta_{\alpha}  + \Delta_3  - \Delta_4\Big)
\langle V_{\alpha}(0)\  V_3(1)V_4(\infty) \rangle,}
\label{l-1l-1C}
\ee
In two iterations of (\ref{virinsC}) and (\ref{w-1C}):
\be
\boxed{
\langle (L_{-1}W_{-1} V_{\alpha})(0)\ V_3(1)V_4(\infty) \rangle}\
\stackrel{(\ref{virinsC})}{=}\
\Big((\Delta_{\alpha}+1)  + \Delta_3  - \Delta_4\Big)
\langle W_{-1}V_{\alpha}(0)\  V_3(1)V_4(\infty) \rangle \\
\boxed{ \stackrel{(\ref{w-1C})}{=}\
\Big(\Delta_{\alpha}  + \Delta_3  - \Delta_4+1\Big)
\Big(\left(w_\alpha + w_3+w_4\right)
\langle V_{\alpha}(0)\ V_3(1)\ V_4(\infty) \rangle \
+ \underline{\langle V_{\alpha}(0)\ (W_{-1}V_3)(1)\ V_4(\infty)
\rangle}\Big)
}
\label{l-1w-1C}
\ee
Again in the last equality we used
$\Delta(W_{-1}V_\alpha) = \Delta_\alpha + 1$.
Underlined are the $\Gamma$-vertices in the answers, which are different from
$\langle V_{\alpha}(0)\ V_1(1)\ V_2(0) \rangle$.

\subsubsection{The difficult case: $W_{-1}^2$}

For three primaries $V_\alpha$, $V_1$, $W_2$
and for $V_{\check\alpha} = W_{-1}V_\alpha$
it follows immediately from (\ref{Wcor3npC}):
\be
\boxed{
\langle (W^2_{-1}V_{\alpha})(0)\ V_3(1)\ V_4(0) \rangle\ = } \\
\stackrel{(\ref{Wcor3npC})}{=}\
\langle (W_{0}W_{-1}V_{\alpha})(0)\ V_3(1)\ V_4(0) \rangle \
+ \langle (W_{-1}V_{\alpha})(0)\ (W_{0}V_3)(1)\ V_4(0) \rangle \
+ \langle (W_{-1}V_{\alpha})(0)\ V_3(1)\ (W_{0}V_4)(0) \rangle \ +
\\
+  \langle (W_{-1}V_{\alpha})(0)\ (W_{-1}V_3)(1)\ V_4(0) \rangle \
- \ \langle (W_{-1}V_{\alpha})(0)\ V_3(1)\ (W_{1}V_4)(0) \rangle\ = \\
= \langle (W_{0}W_{-1}V_{\alpha})(0)\ V_3(1)\ V_4(0) \rangle \
+ (w_3+w_4)\langle (W_{-1}V_{\alpha})(0)\ V_3(1)\ V_4(0) \rangle \
+  \langle (W_{-1}V_{\alpha})(0)\ (W_{-1}V_3)(1)\ V_4(0) \rangle \
\label{Wcor3np11}
\ee
The first term at the r.h.s. can be handled with the help of
(\ref{WWpri}), while (\ref{w-1C}) is directly applicable to the second term .
To handle the last term,
one applies the same (\ref{Wcor3npC})
but with non-primary $V_1\rightarrow W_{-1}V_1$ and
primary $V_{\check\alpha} = V_\alpha$:
\be
\langle (W_{-1}V_{\alpha})(0)\ (W_{-1}V_3)(1)\ V_4(0) \rangle \ =\\
\stackrel{(\ref{Wcor3npC})}{=}\
\langle W_{0}V_{\alpha}(0)\ W_{-1}V_3(1)\ V_4(0) \rangle \
+ \langle V_{\alpha}(0)\ (W_{0}W_{-1}V_3)(1)\ V_4(0) \rangle \
+ \langle V_{\alpha}(0)\ W_{-1}V_3(1)\ (W_{0}V_4)(0) \rangle \ +
\\
+  \langle V_{\alpha}(0)\ (W^2_{-1}V_3)(1)\ V_4(0) \rangle \
- \ \langle V_{\alpha}(0)\ V_3(1)\ (W_{1}V_4)(0) \rangle\ =\\
= (w_\alpha+w_4)\langle  W_{0}V_{\alpha}(0)\ W_{-1}V_3(1)\ V_4(0) \rangle \
+ \langle V_{\alpha}(0)\ (W_{0}W_{-1}V_3)(1)\ V_4(0) \rangle \
+  \langle V_{\alpha}(0)\ (W^2_{-1}V_3)(1)\ V_4(0) \rangle \
\label{Wcor3np1-1}
\ee
Note that there is one less term  at the r.h.s.
than in (\ref{barWcor1np1-1}).

The next step is to substitute
\be
W_0W_{-1} \longrightarrow w W_{-1} + \frac{9D}{2}L_{-1},
\ \ \ \ \ \ \ \ \ \
W_1W_{-1} \longrightarrow \frac{9D}{2}L_0
\ee
from (\ref{WWpri}).
Then one gets from (\ref{Wcor3np11})
with substituted (\ref{Wcor3np1-1}):
\be
\langle (W^2_{-1}V_{\alpha})(0)\ V_3(1)\ V_4(0) \rangle\
= \frac{9D_\alpha}{2}\langle L_{-1}V_{\alpha}(0)\ V_3(1)\ V_4(0) \rangle \
+ (w_\alpha+w_3+w_4)\langle (W_{-1}V_{\alpha})(0)\ V_3(1)\ V_4(0) \rangle +
\\
+(w_\alpha+w_3+w_4)\langle  V_{\alpha}(0)\ W_{-1}V_3(1)\ V_4(0) \rangle
+\! \frac{9D_3}{2}\langle V_{\alpha}(0)\ (L_{-1}V_3)(1)\ V_4(0) \rangle
+  \langle V_{\alpha}(0)\ (W^2_{-1}V_3)(1)\ V_4(0) \rangle
\ee
It remains to substitute (\ref{w-1C}), (\ref{l-1C}) and (\ref{L-1VC})
and get:
\be
\langle (W^2_{-1}V_{\alpha})(0)\ V_3(1)\ V_4(0) \rangle\ = \\ =
(w_\alpha+w_3+w_4)\Big(
\left(w_\alpha + w_3+w_4\right)
\langle V_{\alpha}(0)\ V_3(1)\ V_4(\infty) \rangle \
+ 2\underline{\langle V_{\alpha}(0)\ (W_{-1}V_3)(1)\ V_4(\infty)\rangle}
\Big) + \\
+ \frac{9D_\alpha}{2}
\Big(\Delta_{\alpha}  + \Delta_3  - \Delta_4\Big)
\langle V_{\alpha}(0)\  V_3(1)V_4(\infty) \rangle
- \frac{9D_3}{2}\Big(\Delta_{\check\alpha} +\Delta_3-\Delta_4\Big)
\langle V_{\check\alpha})(0)\ V_3(1)\ V_4(\infty)\rangle + \\
+  \underline{\underline{
\langle V_{\alpha}(0)\ (W^2_{-1}V_3)(1)\ V_4(0) \rangle}}
\label{w-1w-1C}
\ee

\subsection{Restriction to special state}

If $V_1$ and $V_3$ are {\it special} states
($W$-null vectors at the first level), which satisfy
\be
W_{-1} V_1 = \frac{3w_1}{2\Delta_1} L_{-1} V_1, \ \ \ \ \ \ \
W_{-1} V_3 = \frac{3w_3}{2\Delta_3} L_{-1} V_3
\label{spec}
\ee
and
\be
D_1\Delta_1^2 = w_1^2,\ \ \ \ \ \ D_3\Delta_3^2 = w_3^2
\label{sped}
\ee
then underlined correlators are also expressed through
$\langle V_{\alpha})|V_1(1)\ V_2(0)\rangle$ and
$\langle V_{\alpha})(0)\ V_3(1)\ V_4(\infty)\rangle$.

\subsubsection{$\bar\Gamma$-type vertices}

Directly from the definition of the special state
\be
\underline{\langle V_{\alpha} | (W_{-1}V_1)(1)\ V_2(0)\rangle}
\ \stackrel{(\ref{spec})}{=}\ \frac{3w_1}{2\Delta_1}
\langle V_{\alpha} | (L_{-1}V_1)(1)\ V_2(0)\rangle
\stackrel{(\ref{L-1V1C})}{=}\
\frac{3w_1}{2\Delta_1}
\Big(\Delta_{\alpha} - \Delta_1-\Delta_2\Big)
\langle V_{\alpha} | V_1(1) V_2(0)\rangle
\label{specW-1}
\ee
This expression is easy to substitute into (\ref{W-1C}),
(\ref{W-2C}) and (\ref{L-1W-1C}).

In handling the doubly-underlined vertex
in (\ref{W-1W-1C}), one needs also
to use the commutation relation
\be
\left[W_{-1},L_{-1}\right] = W_{-2}
\label{WLcre}
\ee
and the fact that (\ref{L-1V1C}) remains true for
the first descendant $W_{-1}V_1$.
Taking both these things into account, one obtains
\be
\underline{\underline{
\langle V_{\alpha} | (W_{-1}^2V_1)(1)\ V_2(0) \rangle}}
\ \stackrel{(\ref{spec})}{=}\ \frac{3w_1}{2\Delta_1}
\langle V_{\alpha} | (W_{-1}L_{-1}V_1)(1)\ V_2(0) \rangle
\ \stackrel{(\ref{WLcre})}{=}\ \frac{3w_1}{2\Delta_1}
\Big(\langle V_{\alpha} | (L_{-1}W_{-1}V_1)(1)\ V_2(0) \rangle +\\
+ \langle V_{\alpha} | (W_{-2}V_1)(1)\ V_2(0)\rangle\Big)
\ \stackrel{(\ref{L-1V1C})\&(\ref{AABbarC})}{=}\ \frac{3w_1}{2\Delta_1}
\Big(\Delta_{\alpha} - (\Delta_1+1)-\Delta_2 \Big)
\underline{\langle V_{\alpha} | (W_{-1}V_1)(1)\ V_2(0)\rangle} +
\\ +
\frac{3w_1}{2\Delta_1}\Big(\Big(w_{\alpha}-w_1-w_2\Big)
\langle V_{\alpha}|V_1(1)\ V_2(0) \rangle\ -
2\underline{\langle V_{\alpha}|(W_{-1}V_1)(1)\ V_2(0) \rangle}\Big)
= \\  \stackrel{(\ref{specW-1})}{=}\
\frac{3w_1}{2\Delta_1}\left(\Big(w_{\alpha}-w_1-w_2\Big)
+ \Big(\Delta_{\alpha} -\Delta_1-\Delta_2-3 \Big)
\frac{3w_1}{2\Delta_1}
\Big(\Delta_{\alpha} - \Delta_1-\Delta_2\Big)
\right)
\langle V_{\alpha}|V_1(1)\ V_2(0) \rangle
\ee
Substitution into (\ref{W-1W-1C}) gives
\be
\boxed{
\langle W_{-1}^2V_{\alpha} | V_1(1)\ V_2(0) \rangle}\ = \
\left\{(w_\alpha+ 2w_1-w_2)\left(
(w_\alpha+2w_1-w_2)
+ 2\frac{3w_1}{2\Delta_1}
\Big(\Delta_{\alpha} - \Delta_1-\Delta_2\Big)
\right) + \right.
\\
+ \frac{9D_\alpha}{2}
\Big(\Delta_{\alpha}  + \Delta_1  - \Delta_2\Big)
+ 9D_1
\Big(\Delta_{\alpha} - \Delta_1-\Delta_2\Big)
+ \frac{9D_1\Delta_1}{2} +
\\
\left.  +
\frac{3w_1}{2\Delta_1}\left(\Big(w_{\alpha}-w_1-w_2\Big)
+ \Big(\Delta_{\alpha} -\Delta_1-\Delta_2-3 \Big)
\frac{3w_1}{2\Delta_1}
\Big(\Delta_{\alpha} - \Delta_1-\Delta_2\Big)
\right)\right\}
\langle V_{\alpha}|V_1(1)\ V_2(0) \rangle\ = \nn\\
\stackrel{(\ref{sped})}{=}\
\left\{(w_\alpha+ 2w_1-w_2)^2 + \frac{3w_1}{2\Delta_1}
\Big(2(w_\alpha+ 2w_1-w_2)(\Delta_{\alpha} - \Delta_1-\Delta_2)
+(w_{\alpha}-w_1-w_2) \underline{+3w_1}  \Big)
+ \right.\nn\\  \left.
+ \left(\frac{3w_1}{2\Delta_1}\right)^2
\Big(\Delta_{\alpha} -\Delta_1-\Delta_2-3 \underline{+4}\Big)
\Big(\Delta_{\alpha} - \Delta_1-\Delta_2\Big)
+ \frac{9D_\alpha}{2}\Big(\Delta_{\alpha}+ \Delta_1- \Delta_2\Big)
\right\}\langle V_{\alpha}|V_1(1)\ V_2(0) \rangle\ =
\ee
\fr{=
\left\{ \left( w_\alpha+ 2w_1-w_2 + \frac{3w_1}{2\Delta_1}
\Big(\Delta_{\alpha} - \Delta_1-\Delta_2\Big) \right)
\left( w_\alpha+ 2w_1-w_2 + \frac{3w_1}{2\Delta_1}
\Big(\Delta_{\alpha} - \Delta_1-\Delta_2+1\Big) \right)
+ \right. \\ \left.
+ \frac{9D_\alpha}{2}\Big(\Delta_{\alpha}+ \Delta_1- \Delta_2\Big)
\right\}\langle V_{\alpha}|V_1(1)\ V_2(0) \rangle
\label{W-1W-1Cs}}
Underlined are the two terms, obtained when $D_1$ is expressed
through $w_1$ and $\Delta_1$ with the help of (\ref{sped}).

\subsubsection{$\Gamma$-type vertices}

Directly from the definition (\ref{spec}) of the special state
\be\hspace{-.7cm}
\underline{\langle V_{\check\alpha}(0)\ (W_{-1}V_3)(1)\ V_4(\infty)\rangle}\
\stackrel{(\ref{spec})}{=}\ \frac{3w_3}{2\Delta_3}
\langle V_{\check\alpha}(0)\ (L_{-1}V_3)(1)\ V_4(\infty)\rangle\
\stackrel{(\ref{L-1VC})}{=}\
-\frac{3w_3}{2\Delta_3}\Big(\Delta_{\check\alpha} +\Delta_3-\Delta_4\Big)
\langle V_{\check\alpha})(0)\ V_3(1)\ V_4(\infty)\rangle
\label{specw-1}
\ee
This expression is easy to substitute into (\ref{w-1C}),
(\ref{w-2C}) and (\ref{l-1w-1C}).

In handling the doubly-underlined vertex
in (\ref{w-1w-1C}), one needs also
to use the commutation relation (\ref{WLcre})
and the fact that (\ref{L-1VC}) remains true for
the first descendant $W_{-1}V_1$.
Taking both these things into account, one obtains
\be
\underline{\underline{
\langle V_{\alpha}(0)\ (W^2_{-1}V_3)(1)\ V_4(0) \rangle}}
\ \stackrel{(\ref{spec})}{=}\ \frac{3w_3}{2\Delta_3}
\langle V_{\alpha}(0)\ (W_{-1}L_{-1}V_3)(1)\ V_4(0) \rangle
\ \stackrel{(\ref{WLcre})}{=}\ \frac{3w_3}{2\Delta_3}
\Big(\langle V_{\alpha}(0)\ (L_{-1}W_{-1}V_3)(1)\ V_4(0) \rangle +\\
+\ \langle V_{\alpha}(0)\ (W_{-2}V_3)(1)\ V_4(0) \rangle\Big)
\ \stackrel{(\ref{L-1VC})\&(\ref{AABC})}{=}\ -\frac{3w_3}{2\Delta_3}
\Big(\Delta_\alpha+(\Delta_3+1)-\Delta_4\Big)
\underline{\langle V_{\alpha}(0)\ (W_{-1}V_3)(1)\ V_4(0) \rangle\Big)} -
\\ -
\frac{3w_3}{2\Delta_3}
\left(\Big(\hat w_{\check\alpha}+w_3+w_4\Big)
\langle V_{\check\alpha}(0)\ V_3(1)V_4(\infty) \rangle\
+ 2\underline{\langle V_{\check\alpha}(0)\
(W_{-1}V_3)(1)V_4(\infty) \rangle}\right)
= \\
\stackrel{(\ref{specw-1})}{=}\
\frac{3w_3}{2\Delta_3}\left(-\Big(\hat w_{\check\alpha}+w_3+w_4\Big)
+ \Big(\Delta_\alpha+\Delta_3-\Delta_4+3\Big)
\frac{3w_3}{2\Delta_3}\Big(\Delta_{\check\alpha} +\Delta_3-\Delta_4\Big)
\right)\langle V_{\check\alpha}(0)\ V_3(1)V_4(\infty) \rangle
\ee
Substitution into (\ref{w-1w-1C}) gives
\be
\boxed{
\langle (W^2_{-1}V_{\alpha})(0)\ V_3(1)\ V_4(0) \rangle}\ =
\left\{ (w_\alpha+w_3+w_4)\Big(
\left(w_\alpha + w_3+w_4\right)
- 2\frac{3w_3}{2\Delta_3}\Big(\Delta_{\check\alpha} +\Delta_3-\Delta_4\Big)
\right. \nn \\ \left.
+ \frac{9D_\alpha}{2}
\Big(\Delta_{\alpha}  + \Delta_3  - \Delta_4\Big)
- \frac{9D_3}{2}\Big(\Delta_{\check\alpha} +\Delta_3-\Delta_4\Big) +
\right. \nn \\ \left.
+ \frac{3w_3}{2\Delta_3}\left(-\Big(w_{\alpha}+w_3+w_4\Big)
+ \Big(\Delta_\alpha+\Delta_3-\Delta_4+3\Big)
\frac{3w_3}{2\Delta_3}\Big(\Delta_{\check\alpha} +\Delta_3-\Delta_4\Big)
\right)
\right\}\langle V_{\alpha}(0)\ V_3(1)\ V_4(\infty) \rangle = \nn\\
\stackrel{(\ref{sped})}{=}\
\left\{(w_\alpha+w_3+w_4)^2 -
\frac{3w_3}{2\Delta_3}\Big(
2(w_\alpha+w_3+w_4)(\Delta_{\alpha} +\Delta_3-\Delta_4)
+ (w_{\alpha}+w_3+w_4)
\Big) +
\right. \nn \\ \left. +
\left(\frac{3w_3}{2\Delta_3}\right)^2
\Big(\Delta_\alpha+\Delta_3-\Delta_4+3 \underline{-2}\Big)
\Big(\Delta_{\check\alpha} +\Delta_3-\Delta_4\Big)
 + \frac{9D_\alpha}{2}
\Big(\Delta_{\alpha}  + \Delta_3  - \Delta_4\Big)
\right\}\langle V_{\alpha}(0)\ V_3(1)\ V_4(\infty) \rangle\ =
\ee
\fr{
= \left\{\left(w_\alpha+w_3+w_4  -
\frac{3w_3}{2\Delta_3}\Big(\Delta_{\alpha} +\Delta_3-\Delta_4\Big)\right)
\left(w_\alpha+w_3+w_4  -
\frac{3w_3}{2\Delta_3}\Big(\Delta_{\alpha} +\Delta_3-\Delta_4+1\Big)\right)
+   \right.\\
\left.
+\frac{9D_\alpha}{2}
\Big(\Delta_{\alpha}  + \Delta_3  - \Delta_4\Big)
\right\} \langle V_{\alpha}(0)\ V_3(1)\ V_4(\infty) \rangle
\label{w-1w-1Cs}}

\subsection{The free-field-model test}

One can now perform a final test of these formulas, by checking
if (\ref{frefco}) and (\ref{expoexpanLW}) match for this restricted
set of states, with the special values of $\vec\alpha_1\vec\alpha_3$.
For this one should combine our expressions for $\bar\Gamma$,
$Q$ and $\Gamma$ and substituted particular {\it special} values
for $\vec\alpha_1$ and $\vec\alpha_3$. There are $6$ special values
for $\vec\alpha_1$ and $6$ for $\vec\alpha_3$, defined as
zeroes of the Kac determinant $\det Q$ and they are
explicitly given by
\be
\vec\alpha_{spec} = (\alpha, Q\pm Q/2) \ \ \ {\rm or}\ \ \
(\alpha, Q\pm(\alpha\sqrt{3} \pm Q)
\ee
with arbitrary $\alpha$.
Thus, for the $36$ combination of special values
one gets $36$ possibilities for $\vec\alpha_1\vec\alpha_3$ to appear
in (\ref{frefco}). If $Q=0$ there are less: only three possibilities
$\vec\alpha_1\vec\alpha_3 = \alpha_1\alpha_3,\ 4\alpha_1\alpha_3,\
-2\alpha_1\alpha_3$.
Actually, for $Q\neq 0$ also not all of the $6$ values correspond to the
{\it special} vectors: the condition
\be
0 = \left(W_{-1}-\frac{3w_\alpha}{2\Delta_\alpha}L_{-1}\right)V_\alpha
\ \ \  \stackrel{(\ref{L-expQ})\&(\ref{W-expQ})}{=}\\ =
\ \no 3\left(\Big(\alpha^2-\beta^2+\frac{1}{2}Q\beta - \alpha\frac{w}{2\Delta}
\Big)\p\phi_1
+ \Big(-2\alpha\beta + \frac{3}{2}Q\alpha - \beta\frac{w}{2\Delta} \Big)
\p\phi_2\right)e^{\alpha\phi_1+\beta\phi_2}\no  \ =\\ =
{3\over 2\Delta}\Big(\beta-{Q\over 2}\Big)\Big(3\alpha^2-\beta^2\Big)\no \Big(
(\beta-2Q)
\p\phi_1
- \alpha
\p\phi_2\Big)e^{\alpha\phi_1+\beta\phi_2}\no
\ee
is more restrictive:
\be\label{ss}
\vec\alpha_{spec} = (\alpha, Q/2) \ \ \ {\rm or}\ \ \
(\alpha, \pm\alpha\sqrt{3})
\ee
leaving only half of the states, and only such additionally restricted special
values should be included into the checks for $Q\neq 0$.
Evaluation of (\ref{expoexpanLW}) is a little more involved.
However, it nicely reproduces (\ref{frefco}) in all these
$9$ cases, for arbitrary values of $Q$.
This does not prove, but tests the recursion relations,
and the test is positive.

\subsection{Alternative choice of special states}

It is also instructive to look at the case when
special are states $V_2$ and $V_4$. At level one,
we can use the
implications of (\ref{barWcor2}) and (\ref{Wcor4}),
\be
\langle W_{-1}V_\alpha| V_1(1) V_2(0) \rangle>\
\stackrel{(\ref{barWcor2})}{=}\
(2w_\alpha + w_1-2w_2)\langle V_\alpha| V_1(1) V_2(0) \rangle> +
\langle V_\alpha| V_1(1) W_{-1}V_2(0) \rangle>\
\stackrel{(\ref{spec})}{=}\ \\ =
(2w_\alpha + w_1-2w_2)\langle V_\alpha| V_1(1) V_2(0) \rangle> +
\frac{3w_2}{2\Delta_2}\langle V_\alpha| V_1(1) L_{-1}V_2(0) \rangle>\
\stackrel{(\ref{barGrelVfull})}{=}\\ =
(2w_\alpha + w_1-2w_2)\langle V_\alpha| V_1(1) V_2(0) \rangle>
-\frac{3w_2}{2\Delta_2}\langle V_\alpha| V_1(1) L_{-1}V_2(0) \rangle>\
\stackrel{(\ref{L-1V1})}{=}\\
= \left(2w_\alpha + w_1-2w_2  -\frac{3w_2}{2\Delta_2}
(\Delta_\alpha-\Delta_1-\Delta_2)\right)
\langle V_\alpha| V_1(1) V_2(0) \rangle>
\ee
and
\be
\langle (W_{-1}V_\alpha)(0)\ V_3(1) V_4(\infty) \rangle>\
\stackrel{(\ref{Wcor4})}{=}\
(2w_\alpha-w_3+w_4)\langle V_\alpha(0) V_3(1) V_4(\infty) \rangle>
-\langle V_\alpha(0)\ V_3(1)\ (W_{-1}V_4)(\infty) \rangle>\
\stackrel{(\ref{spec})}{=}\\
=(2w_\alpha-w_3+w_4)\langle V_\alpha(0) V_3(1) V_4(\infty) \rangle>
-\frac{3w_4}{2\Delta_4}
\langle V_\alpha(0) V_3(1)\ (L_{-1}V_4)(\infty) \rangle>\
\stackrel{(\ref{Grel2L})\&(\ref{L-1V3})}{=}\\
=\left(2w_\alpha-w_3+w_4+\frac{3w_4}{2\Delta_4}
(\Delta_\alpha-\Delta_3-\Delta_4)\right)
\langle V_\alpha(0) V_3(1) V_4(\infty) \rangle>
\ee
to get
\be
\frac{9D_\alpha\Delta_\alpha}{2}(\Delta_\alpha+\Delta_1-\Delta_2)
(\Delta_\alpha+\Delta_3-\Delta_4)
-3w_\alpha(\Delta_\alpha+\Delta_1-\Delta_2)
\left(2w_\alpha-w_3+w_4+\frac{3w_4}{2\Delta_4}
(\Delta_\alpha-\Delta_3-\Delta_4)\right) -\\
-3w_\alpha \left(2w_\alpha + w_1-2w_2  -\frac{3w_2}{2\Delta_2}
(\Delta_\alpha-\Delta_1-\Delta_2)\right)
(\Delta_\alpha+\Delta_3-\Delta_4) + \\
+2\Delta_\alpha
\left(2w_\alpha + w_1-2w_2  -\frac{3w_2}{2\Delta_2}
(\Delta_\alpha-\Delta_1-\Delta_2)\right)
\left(2w_\alpha-w_3+w_4+\frac{3w_4}{2\Delta_4}
(\Delta_\alpha-\Delta_3-\Delta_4)\right) =
\\
= -18(D_\alpha\Delta^2_\alpha - w_\alpha^2)
(\alpha_1\alpha_3+\beta_1\beta_3)
= 9(D_\alpha\Delta^2_\alpha - w_\alpha^2)
(-2\vec\alpha_1\vec\alpha_3),
\label{match24}
\ee
provided the free-field-theory selection rules
are imposed in addition to the speciality constraints
like (\ref{spec}) on $\vec\alpha_2$ and $\vec\alpha_4$.
Eq.(\ref{match24}) is
exactly what is required by the matching condition between
(\ref{expoexpanLW}) and (\ref{frefco}).

\section{Conclusion}

In this paper we derived and then validated with the help of
the free field model a number of formulas for universal
parts of the triple functions for the Virasoro and $W^{(3)}$
algebras. The Virasoro formulas are elementary and well-known,
their generalization to the $W$-algebra is more tedious and
the answers are less elegant. These formulas are important,
among other things, for study of the AGT relations between
conformal blocks and Nekrasov functions, and we listed
a few concrete expressions that are used in \cite{MMMagt}
and \cite{mmAGT} for this purpose.
As mentioned in the Introduction, if considered at the level
of original Nekrasov integrals, the AGT relations look very
natural in conformal theory, especially from the point of view
of the free field approach.
They will hopefully refresh interest to semi-abandoned
studies of conformal blocks and $W$-symmetries.
Among other things, this technical review can serve as an introduction
to the CFT part of the subject, which emphasizes some
aspects of the theory which are important for AGT studies,
but not well represented in existing CFT literature.

\section*{Acknowledgements}

We are indebted for hospitality and support to Prof.T.Tomaras and
the Institute of Theoretical and Computational Physics of
University of Crete, where part of this work was done.

The work was partly supported by Russian Federal Nuclear Energy
Agency and by RFBR grants 07-02-00878 (A.Mir.),
08-02-00287 (S.Mir.), 07-02-00645 (Al.Mor.) and 07-01-00526 (An.Mor.).
The work of A.Mir. and Al.Mor.
was partly supported by grant FP7-REGPOT-1 (Crete HEP Cosmo 228644),
by joint grants 09-02-90493-Ukr,
09-02-93105-CNRSL, 09-01-92440-CE, 09-02-91005-ANF, INTERREG IIIA
(Greece-Cyprus) and by Russian President's Grant
of Support for the Scientific Schools NSh-3035.2008.2. The work of S.Mir. and of
An.Mor. was partly supported by Russian President's Grant
of Support for the Scientific Schools NSh-3036.2008.2.

\end{document}